\input phyzzx.tex

\Pubnum={\cr
          OU-TAP 31}

\baselineskip 8mm

\leftline{\baselineskip20pt\vbox to10pt
    {\fourteensl\hbox{{~~~~Osaka  University}}
    {\fourteensl\hbox{{Theoretical Astrophysics}} } } }

\def\OSK1{\address{$^1$ Department of Earth and Space Science, Faculty of 
Science, \break Osaka University,~Toyonaka, Osaka 560,~Japan}}

\def\THK{\address{$^2$ Astronomical Institute, Graduate School of 
Science, \break Tohoku University,~Sendai 980-77,~Japan}}

\newdimen\HOFFSET  \HOFFSET=-35pt

\titlepage

\hsize 17.2cm
\vsize 24cm

\vskip 5mm

\title{\bf Post-Newtonian Hydrodynamic Equations Using the (3+1) Formalism 
in General Relativity} 

\author{Hideki Asada$^1$, Masaru Shibata$^1$ and Toshifumi Futamase$^2$}
\OSK1

\THK

\abstract{
Using the (3+1) formalism in general relativity, 
we perform the post-Newtonian(PN) approximation 
to clarify what sort of gauge condition is suitable for 
numerical analysis of coalescing compact binary neutron stars and 
gravitational waves from them. 
We adopt a kind of transverse gauge condition to determine the shift vector. 
On the other hand, for determination of the time slice, 
we adopt three slice conditions(conformal 
slice, maximal slice and harmonic slice) and discuss their properties. 
Using these conditions, the PN hydrodynamic equations are obtained 
up through the 2.5PN order including the quadrupole gravitational radiation 
reaction. In particular, we describe methods to solve 
the 2PN tensor potential which arises from the spatial 3-metric.
It is found that the conformal slice seems appropriate for analysis of 
gravitational waves in the wave zone and the maximal slice will be useful 
for describing the equilibrium configurations. 
The PN approximation in the (3+1) formalism 
will be also useful to perform numerical simulations using various 
slice conditions and, as a result, to provide an initial data 
for the final merging phase of coalescing binary neutron stars which 
can be treated only by fully general relativistic simulations. 
}


\REF\ligo{A. Abramovici et al. Science, {\bf 256}(1992), 325; \nextline
K. S. Thorne, in proceedings of the eighth Nishinomiya-Yukawa
memorial symposium on Relativistic Cosmology, edited by
M. Sasaki(Universal Academy Press, Tokyo, 1994), pp.67.}

\REF\virgo{C. Bradaschia et al., Nucl. Instrum. Method Phys. Res. Sect.
{\bf A289}(1990), 518.}

\REF\phinney{E. S. Phinney, Astrophys. J. Lett. {\bf 380}(1991), 17.}

\REF\will{For example, C. M. Will, in proceedings of the eighth 
Nishinomiya-Yukawa memorial symposium on Relativistic Cosmology(ref.1), 
pp.83; See also references cited therein. }

\REF\blan{L. Blanchet et al., Phys. Rev. Lett. {\bf 74}(1995), 3515;
as for detailed description of their calculations, see references cited 
therein :
\nextline L. E. Kidder, Phys. Rev. {\bf D52}(1995), 821. }

\REF\blanchet{L. Blanchet: He has already completed the  
energy luminosity accurate up to 2.5PN order(without spin dependent terms)
extending the work by Blanchet et al.(ref.\blan).}

\REF\tagoshi{H. Tagoshi and T. Nakamura, Phys. Rev. {\bf D49}(1994), 4016
; \nextline
H. Tagoshi and M. Sasaki, Prog. Theor. Phys. {\bf 92}(1994), 745; 
\nextline M. Shibata, M. Sasaki, H. Tagoshi and T. Tanaka, Phys. Rev. 
{\bf D51}(1995), 1646; \nextline
H. Tagoshi, M. Shibata, T. Tanaka and M. Sasaki, submitted to Phys. Rev. D.}

\REF\cult{C. Cutler, Phys. Rev. Lett. {\bf 70}(1993), 2984.}

\REF\cutler{C. Cutler and E. E. Flanagan, Phys. Rev. {\bf D49}(1994), 2658.}

\REF\mark{B. F. Schutz, Nature, {\bf 323}(1986), 210; \nextline 
D. Markovic, Phys. Rev. {\bf D48}(1993), 4738.}

\REF\lai{D. Lai, F. Rasio and S. L. Shapiro, Astrophys. J. {\bf 420}
(1994), 811; Astrophys. J. supplement {\bf 88}(1993), 205.}

\REF\kidder{L. E. Kidder, C. M. Will and A. G. Wiseman, Phys. Rev. 
{\bf D47}(1993), 3281.}

\REF\nakamura{T. Nakamura, in the proceedings of the eighth
Nishinomiya-Yukawa memorial symposium on Relativistic Cosmology(ref.1), 
pp.155.}

\REF\oohara{K. Oohara and T. Nakamura, Prog. Theor. Phys. {\bf 83}(1990), 
906; {\bf 86}(1991), 73; {\bf 88}(1992), 307.}

\REF\shibata{M. Shibata, T. Nakamura and K. Oohara, Prog. Theor. Phys. 
{\bf 88}(1992), 1079; {\bf 89}(1993), 809.}

\REF\bds{L. Blanchet, T. Damour and G. Sch\"afer,
Mon. Not. R. Astr. Soc. {\bf 242}(1990), 289.}

\REF\ruffert{M. Reffert, H-T. Janka and G. Sch\"afer, Max-Planck-Institute 
preprint.}

\REF\conf{M. Shibata and T. Nakamura, Prog. Theor. Phys. {\bf 88}(1992), 317.}

\REF\schafer{G. Sch\"afer, Ann. Phys. {\bf 161}(1985), 81.}

\REF\masso{C. Bona and J. Masso, Phys. Rev. Lett. {\bf 68}(1992), 1097.}

\REF\chris{Christodoulou, Phys. Rev. Lett. {\bf  67}(1991), 1486 ; \nextline
A. G. Wiseman and C. M. Will, Phys. Rev. {\bf 44}(1991), 2945 ; \nextline
K. S. Thorne, Phys. Rev. {\bf 45}(1992), 520.}

\REF\chand{S. Chandrasekhar, Astrophys. J. {\bf 142}(1965), 1488.}

\REF\ellp{S. Chandrasekhar, Ellipsoidal Figures of Equilibrium(Dover, 1969).}

\REF\chandra{S. Chandrasekhar and F. P. Esposito, Astrophys. J. {\bf 160}(1970), 55.}

\REF\ohta{T. Ohta. H. Okamura, T. Kimura and K. Hiida,
Prog. Thor. Phys. {\bf 51}(1974), 1598.}

\REF\chandr{S. Chandrasekhar, Astrophys. J. {\bf 158}(1969), 45.}

\REF\wilson{J. R. Wilson and G. J. Mathews, Phys. Rev. Lett. {\bf 75}(1995), 
4161.}

\REF\smarr{L. Smarr and J. W. York, Jr., Phys. Rev. {\bf D17}(1978), 1945; 
2529.}

\def\pa{\partial}
\def\bI{\hbox{$\,I\!\!\!\!-$}}
\def\two{\hbox{$_{(2)}$}}
\def\three{\hbox{$_{(3)}$}}
\def\four{\hbox{$_{(4)}$}}
\def\five{\hbox{$_{(5)}$}}
\def\six{\hbox{$_{(6)}$}}
\def\seven{\hbox{$_{(7)}$}}
\def\eight{\hbox{$_{(8)}$}}

\newdimen\pictureindent           \pictureindent=15pt
\newdimen\str
\newdimen\squareht
\newdimen\squarewd
\newskip\doublecolskip
\newskip\tableoftablesskip        \tableoftablesskip=\baselineskip

\newbox\squarebox

\def\square{%
      \setbox\squarebox=\boxit{\hbox{\phantom{x}}}
      \squareht = 1\ht\squarebox
      \squarewd = 1\wd\squarebox
      \vbox to 0pt{
          \offinterlineskip \kern -.9\squareht
          \hbox{\copy\squarebox \vrule width .0\squarewd height .8\squareht
              depth 0pt \hfill
          }
          \hbox{\kern .2\squarewd\vbox{%
            \hrule height .0\squarewd width \squarewd}
          }
          \vss
      }}

\baselineskip 8mm

\chapter{Introduction}

Kilometer-size interferometric gravitational wave
detectors, such as LIGO\refmark\ligo and VIRGO\refmark\virgo, 
are now in construction aiming at direct
detection of gravitational waves from relativistic astrophysical
objects. Coalescing binary neutron stars are the most 
promising sources of gravitational waves for such detectors. The reasons
are that (1) we expect to detect the signal of coalescence of
binary neutron stars about several times per year\refmark\phinney, and
(2) the wave form from coalescing binaries can be predicted with a high
accuracy compared with other sources\refmark\ligo.

In the case when the orbital separation of each star is 
large compared with the radius of neutron stars, \ie,
in the so-called inspiraling phase,
binary neutron stars are evolving in the adiabatic manner due to 
gravitational radiation reaction with much longer time scale 
than the orbital period.
As for the inspiraling phase, the theoretical investigation is usually done
by the point particle approach using
the PN approximation in general relativity\refmark{\will,
\blan,\blanchet,\tagoshi}.
Since the separation is large compared with the
neutron star radius, the hydrodynamic effect is small enough and we can
regard each star of binary as a point particle.
Theoretical studies for such a phase 
is potentially important because by comparing
the observational signal with the theoretical prediction of the
signal of inspiraling binary, we may be able to know not only the various
parameters of binary\refmark{\cult,\cutler}, but also the cosmological
parameters\refmark\mark.

After a long time emission of gravitational waves,
the orbital separation becomes comparable to the radius of the 
neutron star. Then, each star of binary neutron stars begins to behave
as a hydrodynamic object, not as a point particle,
because they are tidally coupled each other.
Recently, Lai, Rasio and Shapiro\refmark\lai have pointed out that such a
tidal coupling of binary neutron stars is very important for
their evolution in the final merging phase 
because the tidal effect causes the instability to
the circular motion of them.
Also important is the general relativistic gravity because in such a
phase, the orbital separation is larger than $\sim 10\%$ of the
Schwarzschild radius of the system. Thus, we need not only 
a hydrodynamic treatment, but also general relativistic one 
to study the final phase of binary neutron stars.

Fully general relativistic simulation is sure to be the best method,
but it is also one of the most difficult ones. Although much effort has been  
focused and much progress can be expected there\refmark\nakamura, it will 
take a long time until numerical relativistic calculations become 
reliable. One of the main reasons is that 
we do not know the behavior of the geometric variables in the strong field 
around coalescing binary neutron stars. Owing to this, we do not know 
what sort of gauge condition is useful and 
how to give an appropriate general relativistic 
initial condition for coalescing binary neutron stars. 

The other reason is a technical one: In numerical relativistic 
simulations, gravitational waves are generated, and in the 
case of coalescing binary neutron stars, the wavelength is the order of  
$\lambda \sim \pi R^{3/2}M^{-1/2}$, where $R$ and $M$ are the orbital radius 
and the total mass of binary, respectively. Thus,  we need 
to cover a region $L >\lambda \propto R^{3/2}$ with numerical grids 
in order to perform accurate simulations. 
This is in contrast with the case of Newtonian and/or PN
simulations, in which we only need to cover a region $\lambda > L > R$. 
Since the circular orbit of binary neutron stars becomes unstable at 
$R \lsim 10M$ owing to the tidal effects\refmark\lai or the strong 
general relativistic gravity\refmark\kidder, we must set 
an initial condition of binary at $R \gsim 10M$. For such a case, 
to perform an accurate simulation, 
the grid must cover a region $L>\lambda \sim 100M$ in numerical relativistic 
simulations. When we assume to cover each neutron star of its radius $\sim 5M$ 
with $\sim 30$ homogeneous grid\refmark{\oohara,\shibata}, 
we need to take grids of at least $\sim 500^3$, 
but it seems impossible to take such a large amount of mesh points 
for the present power of supercomputer. 
At present, we had better search other methods to prepare an initial 
condition for binary neutron stars. 

In the case of PN simulations, the situation is completely 
different because we do not have to treat gravitational waves 
explicitly in numerical simulations, and as the result, 
only need to cover a region at most $L \sim 20-30M$. 
In this case, it seems that $\sim 200^3$ grid numbers are enough. 
Furthermore, we can take into account general 
relativistic effects with a good accuracy: In the case of coalescing binary 
neutron stars, the error will be at most $\sim M/R \sim {\rm a~few} \times 
10\%$ for the first PN
approximation, and $\sim (M/R)^2 \sim {\rm several}~ \%$ for 2PN 
approximation. 
Hence, if we could take into account up through 2PN terms, we 
would be able to give a highly accurate initial condition(the error 
$\lsim {\rm several}~\%$).
For these reasons, we consider the 2.5PN hydrodynamic equations 
including 2.5PN radiation reaction potential in this paper. 

The purpose of this paper is twofold: One is to establish the basic  
formulation of the 2.5PN hydrodynamic equation, and the other 
is to investigate 
what kind of gauge condition is appropriate for simulation of the coalescing 
binary neutron stars and extraction of gravitational waves from them. 
As for the PN hydrodynamic equation, Blanchet et al. have already 
obtained the (1+2.5)PN formula\refmark\bds.  
Although their formula was very 
useful for PN hydrodynamic simulations including the radiation 
reaction\refmark{\oohara,\shibata,\ruffert}, they did not 
take into account 2PN 
terms. Also, in their formula, they fixed the gauge conditions to the ADM 
gauge, but in numerical relativity, it has not been known yet 
what sort of gauge condition is suitable for simulation of 
the coalescing binary neutron stars and estimation of gravitational waves 
from them. 
For these reasons, we shall investigate several 
gauge conditions using the (3+1) formalism in general relativity.  

This paper is organized as follows. In section 2 we present the (3+1) 
formalism of the Einstein equation 
and the equations for the PN approximation. 
Several slice conditions are imposed in section 3. 
In section 4, the quadrupole radiation-reaction potential is calculated 
in combination of the conformal slice\refmark\conf and the transverse gauge. 
It is found that this combination of the gauge conditions simplifies the 
calculation of the back reaction potential. 
The methods to solve the 2PN tensor potential are discussed  
in detail for the sake of actual numerical simulations in section 5. 
The conserved energy and conserved linear momentum are refered in section 6.
Section 7 is devoted to summary. 

We use the units of $c=G=1$ in this paper. Greek and Latin indices take 
$0,1,2,3$ and $1,2,3$, respectively. 
 
\chapter{(3+1) Formalism for Post-Newtonian Approximation} 

\section{(3+1) Formalism}

We consider the (3+1) formalism to perform the PN 
approximation. In the (3+1) formalism, the metric is split as 
$$
g_{\mu\nu}=\gamma_{\mu\nu}-\hat n_{\mu}\hat n_{\nu},\eqno\eq
$$
and 
$$\eqalign{
&\hat n_{\mu}=(-\alpha,~{\bf 0}),\cr
&\hat n^{\mu}=
\Bigl({1 \over \alpha},~-{\beta^i \over \alpha}\Bigr),\cr}\eqno\eq
$$
where $\alpha, \beta^i$ and $\gamma_{ij}$ are the lapse function, shift 
vector  and metric on the 3D hypersurface, respectively. 
Then the line element is written as 
$$
ds^2=-(\alpha^2-\beta_i \beta^i)dt^2+2\beta_i dt dx^i+
\gamma_{ij}dx^i dx^j. \eqno\eq
$$
Using the (3+1) formalism, the Einstein equation 
$$
G_{\mu\nu}=8\pi T_{\mu\nu},\eqno\eq
$$
is split into the constraint equations and 
the evolution equations. The formers are the so-called Hamiltonian and 
momentum constraints which respectively become 
$$
{\rm tr}R-K_{ij}K^{ij}+K^2=16\pi \rho_{H}, \eqn\hami
$$
$$
D_i K^i_{~j}-D_j K=8\pi J_j, \eqn\cons
$$
where  $K_{ij}$, $K$, ${\rm tr}R$ and $D_i$ are the extrinsic curvature, 
the trace part of $K_{ij}$, the scalar curvature of 3D hypersurface 
and the covariant derivative with respect of $\gamma_{ij}$. 
$\rho_{H}$ and $J_j$ are defined as 
$$\eqalign{
&\rho_{H}=T_{\mu\nu}\hat n^{\mu}\hat n^{\nu},\cr
&J_j=-T_{\mu\nu}\hat n^{\mu}\gamma^{\nu}_{~j}.\cr}\eqno\eq
$$
Evolution equations for the spatial metric and extrinsic curvature 
are respectively 
$$
{\pa \over \pa t} \gamma_{ij}=-2\alpha K_{ij}+D_i \beta_j 
+D_j \beta_i, {\hskip 8.3cm}\eqno\eq
$$
$$\eqalign{
{\pa \over \pa t} K_{ij}&=\alpha(R_{ij}+K K_{ij}-2K_{il}K^l_{~j})
 -D_iD_j \alpha \cr
&+(D_j \beta^m) K_{mi}+(D_i \beta^m) K_{mj}+\beta^m D_m K_{ij} 
-8\pi\alpha\Bigl(S_{ij}+{1 \over 2}\gamma_{ij}(\rho_{H}-S^l_{~l})\Bigr),
\cr}
 \eqno\eq
$$
$$
{\pa \over \pa t} \gamma=2\gamma(-\alpha K+D_i \beta^i), {\hskip 9.5cm}
 \eqno\eq
$$
$$
{\pa \over \pa t} K=\alpha ({\rm tr}R+K^2)-D^iD_i \alpha 
+\beta^j D_j K +4\pi\alpha (S^l_{~l}-3\rho_{H}),{\hskip 3.4cm}
\eqno\eq
$$
where $R_{ij}$, $\gamma$ and $S_{ij}$ are, respectively, 
the Ricci tensor with respect of 
$\gamma_{ij}$, determinant of $\gamma_{ij}$ and 
$$
S_{ij}=T_{kl}\gamma^k_{~i}\gamma^l_{~j}.\eqno\eq
$$
Hereafter we use the conformal factor 
$\psi=\gamma^{1/12}$ instead of $\gamma$ for simplicity. 

To distinguish the wave part from the non-wave part(for example, 
Newtonian potential) in the metric, we use 
$\tilde \gamma_{ij}=\psi^{-4} \gamma_{ij}$ instead of $\gamma_{ij}$. 
Then ${\rm det}(\tilde \gamma_{ij})=1$ is satisfied. 
We also define $\tilde A_{ij}$ as 
$$
\tilde A_{ij} \equiv \psi^{-4} A_{ij}
\equiv \psi^{-4}\Bigl(K_{ij}-{1 \over 3}\gamma_{ij} K\Bigr).\eqno\eq
$$
We should note that in our notation, indices of 
$\tilde A_{ij}$ are raised and lowered by $\tilde \gamma_{ij}$, so that 
the relations, 
$\tilde A^i_{~j}=A^i_{~j}$ and $\tilde A^{ij}=\psi^4A^{ij}$, hold. 
Using these variables, the evolution equations (2.8)-(2.11) can be 
rewritten as follows;
$$
{\pa \over \pa n} \tilde \gamma_{ij}=-2\alpha \tilde A_{ij} 
+\tilde \gamma_{il}{\pa \beta^l \over \pa x^j}
+\tilde \gamma_{jl}{\pa \beta^l \over \pa x^i}
-{2 \over 3}\tilde \gamma_{ij}{\pa \beta^l \over \pa x^l},{\hskip 6.8cm}
\eqn\ggg
$$
$$\eqalign{
{\pa \over \pa n} \tilde A_{ij}=&{1 \over \psi^4} \Bigl[
\alpha \Bigl(R_{ij}-{1 \over 3}\gamma_{ij}{\rm tr}R \Bigr)
-\Bigl(\tilde D_i\tilde D_j \alpha-{1 \over 3}\tilde\gamma_{ij}
\tilde\Delta \alpha \Bigr) 
-{2 \over \psi}\Bigl(\psi_{,i}\alpha_{,j}+\psi_{,j}\alpha_{,i}
-{2 \over 3}\tilde\gamma_{ij}\tilde\gamma^{kl}\psi_{,k}\alpha_{,l} 
\Bigr)\Bigr] \cr
+&\alpha(K \tilde A_{ij}-2\tilde A_{il}\tilde A^l_{~j})
+{\pa \beta^m \over \pa x^i} \tilde A_{mj}
+{\pa \beta^m \over \pa x^j} \tilde A_{mi}
-{2 \over 3}{\pa \beta^m \over \pa x^m} \tilde A_{ij}
-8\pi{\alpha \over \psi^4}
\Bigl(S_{ij}-{1 \over 3}\gamma_{ij}S^l_{~l}\Bigr),\cr}
 \eqn\kkk
$$
$$
{\pa \over \pa n} \psi={\psi \over 6}\Bigl(-\alpha K+
{\pa \beta^i \over \pa x^i}\Bigr), {\hskip 10.6cm} \eqn\kkg
$$
$$
{\pa \over \pa n} K=\alpha \Bigl(\tilde A_{ij}\tilde A^{ij}
+{1 \over 3}K^2 \Bigr) 
-{1 \over \psi^4}\tilde\Delta\alpha-{2 \over \psi^5}\tilde\gamma^{kl} 
\psi_{,k}\alpha_{,l}+4\pi\alpha(S^i_{~i}+\rho_{H}),{\hskip 2.8cm}
\eqn\kkt
$$
where $\tilde D_i$ and $\tilde\Delta$ are 
the covariant derivative and Laplacian with respect to 
$\tilde\gamma_{ij}$ and 
$$
{\pa \over \pa n}={\pa \over \pa t}
-\beta^i{\pa \over \pa x^i}.\eqno\eq
$$
The constraint equations are also written as 
$$
\tilde \Delta \psi={1 \over 8}{\rm tr}\tilde R\psi-2\pi \rho_{H}\psi^5
-{\psi^5 \over 8}\Bigl(\tilde A_{ij} \tilde A^{ij}-{2 \over 3}K^2 \Bigr)
,\eqn\hamisec
$$
and
$$
\tilde D_j (\psi^6 \tilde A^j_{~i})-{2 \over 3}\psi^6 \tilde D_i K=
8 \pi \psi^6 J_i, \eqno\eq
$$
where ${\rm tr}\tilde R$ is the scalar curvature with respect to 
$\tilde\gamma_{ij}$. 

Now let us consider $R_{ij}$ in Eq.$\kkk$, 
which is one of the main source terms of the evolution 
equation for $\tilde A_{ij}$. First we split $R_{ij}$ into two parts as 
$$
R_{ij}=\tilde R_{ij}+R^{\psi}_{ij},\eqno\eq
$$
where $\tilde R_{ij}$ is the Ricci tensor with respect to $\tilde 
\gamma_{ij}$, 
$$
R^{\psi}_{ij}=-{2 \over \psi}\tilde D_i \tilde D_j \psi
-{2 \over \psi} \tilde \gamma_{ij} \tilde D^k \tilde D_k \psi 
+{6 \over \psi^2}(\tilde D_i \psi)(\tilde D_j \psi)
-{2 \over \psi^2}\tilde \gamma_{ij}(\tilde D_k \psi)(\tilde D^k \psi).
 \eqno\eq
$$
Using the property of ${\rm det}(\tilde \gamma_{ij})=1$, 
$\tilde R_{ij}$ is written as 
$$
\tilde R_{ij}={1 \over 2}\Bigl[\tilde \gamma^{kl}
(\tilde \gamma_{lj,ik}+\tilde \gamma_{li,jk}-\tilde \gamma_{ij,lk}) 
+\tilde \gamma^{kl}_{~~,k}
(\tilde \gamma_{lj,i}+\tilde \gamma_{li,j}-\tilde \gamma_{ij,l})\Bigr]
-\tilde \Gamma^l_{kj}\tilde \Gamma^k_{li} ~,  \eqno\eq
$$
where $_{,i}$ denotes $\pa/\pa x^i$ and 
$\tilde \Gamma^k_{ij}$ is the Christoffel symbol with respect to 
$\tilde \gamma_{ij}$. We split $\tilde \gamma_{ij}$ and $\tilde \gamma^{ij}$ 
as $\delta_{ij}+h_{ij}$ and $\delta^{ij}+f^{ij}$, where $\delta_{ij}$ denotes 
the flat metric, and rewrite $\tilde R_{ij}$ as 
$$\eqalign{
\tilde R_{ij}&={1 \over 2}\Bigl[-h_{ij,kk}+h_{jl,li}+h_{il,lj}
+f^{kl}_{~~,k}(h_{lj,i}+h_{li,j}-h_{ij,l})\cr
&~~~~+f^{kl}(h_{kj,il}+h_{ki,jl}-h_{ij,kl})\Bigr]
-\tilde \Gamma^l_{kj} \tilde \Gamma^k_{li}~. \cr}\eqn\ricci
$$

In this paper, we consider only the linear order in $h_{ij}$ and $f_{ij}$ 
assuming $|h_{ij}|,~|f_{ij}| \ll 1$. (So that, $h_{ij}=-f^{ij}$.) 
Such an assumption is justified because in this paper, we choose a gauge 
condition, in which $h_{ij}$ is a 2PN quantity(see below). 
This implies that we neglect higher PN effects such as 
the non-linear coupling between gravitational 
waves themselves, but does not imply that we neglect 
the non-linear coupling between the Newtonian potentials themselves and 
between gravitational waves and the Newtonian potentials. 
In other words, although we can not see the non-linear memory of 
gravitational waves\refmark\chris, we can see the tail term of gravitational 
waves and can derive the exact quadrupole formula(see below). 
Here, to guarantee the wave property of $\tilde \gamma_{ij}$, 
we impose a kind of transverse gauge\footnote{*}{Hereafter, we call this 
condition merely the transverse gauge.} to $h_{ij}$  as
$$
h_{ij,j}=0.\eqno\eq
$$
This condition is guaranteed by $\beta^i$ which satisfies 
$$
-\beta^{k}_{,~j} \tilde \gamma_{ij,k}=\Bigl(-2\alpha \tilde A_{ij} 
+\tilde \gamma_{il} \beta^l_{~,j}+\tilde \gamma_{jl} \beta^l_{~,i}
-{2 \over 3}\tilde \gamma_{ij} \beta^l_{~,l}
 \Bigr)_{,j}.\eqn\bbb
$$
Using the above conditions, Eq. $\ricci$ becomes 
$$
\tilde R_{ij}=-{1 \over 2}\Delta_{flat} h_{ij}+O(h^2),\eqno\eq
$$
where $\Delta_{flat}$ is the Laplacian with respect to $\delta_{ij}$. 
Note that ${\rm tr}\tilde R=O(h^2)$ is guaranteed in the transverse 
gauge because the traceless property of $h_{ij}$ holds 
in the linear order. 

Finally, we show the equations for the perfect fluid. 
The energy momentum tensor for the perfect fluid is written as 
$$
T^{\mu\nu}=(\rho+\rho \varepsilon +P)u^{\mu}u^{\nu}+P g^{\mu\nu},\eqno\eq
$$
where $u^{\mu}$, $\rho$, $\varepsilon$ and $P$ are the four velocity, 
the mass density, the specific internal energy 
and the pressure. The mass density obeys the continuity equation 
$$
\nabla_{\mu} (\rho u^{\mu})=0,\eqn\conti
$$
where $\nabla_{\mu}$ is the covariant derivative with respect to 
$g_{\mu\nu}$. 
The explicit form is 
$$
{\pa \rho_{\ast} \over \pa t}+{\pa (\rho_{\ast}v^i) \over \pa x^i}=0, 
\eqno\eq
$$
where $\rho_{\ast}$ is the conserved density defined as 
$$
\rho_{\ast}=\alpha\psi^6\rho u^0.
\eqno\eq
$$
The equations of motion and the energy equation are derived from 
$$
\nabla_{\mu} T^{\mu\nu}=0.\eqno\eq
$$
Explicit forms of them become 
$$
{\pa S_i \over \pa t}+{\pa (S_i v^j) \over \pa x^j}
=-\alpha \psi^6 P_{,i}-\alpha \alpha_{,i} S^0 + S_j \beta^j_{~,i}
-{1 \over 2S^0}S_j S_k \gamma^{jk}_{~~,i},
\eqn\EOM
$$
and 
$$
{\pa H \over \pa t}+{\pa (H v^j) \over \pa x^j}
=-P\Bigl({\pa (\alpha \psi^6 u^0) \over \pa t}
+{\pa (\alpha \psi^6 u^0 v^j) \over \pa x^j}\Bigr),
\eqn\Energy
$$
where 
$$\eqalign{
&S_i=\alpha \psi^6 (\rho+\rho \varepsilon +P)u^0 u_i
=\rho_{\ast} \Bigl(1+\varepsilon+{P \over \rho} \Bigr)u_i (=\psi^6 J_i),\cr
&S^0=\alpha \psi^6 (\rho+\rho \varepsilon +P)(u^0)^2 \Bigl(=
{(\rho_H +P)\psi^6 \over \alpha}\Bigr),\cr
&H=\alpha \psi^6 \rho \varepsilon u^0=\rho_{\ast}\varepsilon ,\cr
&v^i\equiv {u^i \over u^0}=-\beta^i+{\gamma^{ij} S_j \over S^0}.\cr}\eqn\vel
$$
Finally, we note that in the above equations, only $\beta^i$ appears, 
and $\beta_i$ does not, so that, in the subsequent section, we only 
consider the PN expansion of $\beta^i$, not of $\beta_i$. 

\section{Post-Newtonian approximation}

Next, we consider the PN approximation of 
the above equations. First of all, we review the PN expansion 
of the variables. 
Each metric variable, which is relevant to the present paper, 
is expanded as 
$$\eqalign{
\psi&=1+\two\psi+\four\psi+\six\psi+\seven\psi+\dots,\cr
\alpha&=1+\two\alpha+\four\alpha+\six\alpha+\seven\alpha+\dots,\cr
&=1-U+\Bigl({U^2 \over 2}+X\Bigr)+\six\alpha+\seven\alpha+\dots, \cr
\beta^i&=\three\beta_i+\five\beta_i+\six\beta_i+\seven\beta_i+
\eight\beta_i+\dots,\cr
h_{ij}&=\four h_{ij}+\five h_{ij}+\dots,\cr
\tilde A_{ij}&=\three\tilde A_{ij}+\five\tilde A_{ij}+\six\tilde A_{ij}
+\dots,\cr
K&=\three K+\five K+\six K+\dots,\cr}\eqno\eq
$$
where subscripts denote the PN order($c^{-n}$) and $U$ is 
the Newtonian potential satisfying 
$$
\Delta_{flat} U=-4\pi \rho.\eqn\Npot
$$
$X$ depends on the slice condition, and in the standard PN 
gauge\refmark\chand, we usually use $\Phi=-X/2$, which satisfies 
$$
\Delta_{flat}\Phi=-4\pi\rho\Bigl(v^2+U+{1 \over 2}\varepsilon
+{3 \over 2}{P \over \rho}\Bigr). \eqno\eq
$$
Note that the terms relevant to 
the radiation reaction appear in $\seven\psi$, 
$\seven\alpha$, $\eight\beta_i$ and $\five h_{ij}$, and 
the quadrupole formula is derived from 
$\seven\alpha$ and $\five h_{ij}$.

The four velocity is expanded as 
$$\eqalign{
u^0=&1+\Bigl({1 \over 2}v^2+U\Bigr)+\Bigl({3 \over 8}v^4+{5 \over 2}v^2U
+{1 \over 2}U^2+\three\beta_i v^i-X \Bigr)+O(c^{-6}),\cr
u_0=&-\Bigl[ 1+\Bigl({1 \over 2}v^2-U\Bigr)
+\Bigl({3 \over 8}v^4+{3 \over 2}v^2U
+{1 \over 2}U^2+X \Bigr)\Bigr]+O(c^{-6}),\cr
u^i=&v^i
\Bigl[1+\Bigl({1 \over 2}v^2+U\Bigr)+\Bigl({3 \over 8}v^4+{5 \over 2}v^2U
+{1 \over 2}U^2+\three\beta_i v^i-X \Bigr)\Bigr]+O(c^{-7}),\cr
u_i=&v^i+\Bigl\{\three\beta_i
+v^i\Bigl({1 \over 2}v^2+3 U\Bigr)\Bigr\} 
+\Bigl[\five\beta_i+\three\beta_i\Bigl({1 \over 2}v^2+3U\Bigr)
+\four h_{ij}v^j \cr
&~~~~~+v^i\Bigl({3 \over 8}v^4+{7 \over 2}v^2 U+4U^2-X+4\four\psi 
+\three\beta_j v^j\Bigr)\Bigr]+\Bigl(\six \beta_i+\five h_{ij}v^j\Bigr)
+O(c^{-7}), \cr}
\eqno\eq
$$
where $v^2=v^i v^i$. 
The PN expansion of the relation $u^{\mu}u_{\mu}=-1$ becomes 
$$\eqalign{
(\alpha u^0)^2&=1+\gamma^{ij}u_i u_j\cr
&=1+v^2+v^4+4v^2U+2\three\beta_iv^i+O(c^{-6}).\cr}\eqno\eq
$$
Thus $\rho_{H}$, $J_i$ and $S_{ij}$ are respectively expanded as
$$
\eqalign{
\rho_{H}=&\rho\Bigl[1+\Bigl(v^2+\varepsilon\Bigr)+\Bigl\{v^4+v^2\Bigl(
4U+\varepsilon+{P \over \rho}\Bigr)+2\three\beta_i v^i\Bigr\}
+O(c^{-6})\Bigr], \cr
J_i=&\rho\Bigl[v^i\Bigl(1+v^2+3U+\varepsilon+{P \over \rho}\Bigr)
+\three\beta_i+O(c^{-5})\Bigr], \cr 
S_{ij}=&\rho\Bigl[\Bigl(v^i v^j+{P \over \rho}\delta_{ij}\Bigr) 
+\Bigl\{\Bigl(v^2+6U+\varepsilon+{P \over \rho}\Bigr)v^i v^j
+v^i\three\beta_j+v^j\three\beta_i+2{UP \over \rho}\delta_{ij} \Bigr\} \cr
&~~~~~~~~~~~~~~~~~~~~~~~~~+O(c^{-6})\Bigr], \cr
S_l^{~l}=&\rho\Bigl[v^2+{3P \over \rho}+\Bigl\{2 \three \beta_i v^i+
v^2\Bigl(v^2+4U+\varepsilon+{P \over \rho}\Bigr)\Bigr\}+O(c^{-6})\Bigr]. \cr}
\eqno\eq
$$

$\psi$(and $\alpha$ in the conformal slice) is determined by 
the Hamiltonian constraint. At the lowest order, it becomes 
$$
\Delta_{flat} \two\psi=-2\pi \rho . \eqno\eq
$$
Thus, $\two\alpha=-2\two\psi=-U$ is satisfied in this paper. 
At the 2PN and 3PN orders, the Hamiltonian constraint equation 
becomes, respectively, 
$$
\Delta_{flat} \four\psi=-2\pi\rho\Bigl(v^2+\varepsilon+{5 \over 2}U \Bigr),
\eqn\psifo
$$
and
$$\eqalign{
\Delta_{flat} \six\psi=&
-2\pi\rho\Bigl\{v^4+v^2\Bigl(\varepsilon+{P \over \rho}+{13 \over 2}U
\Bigr)+2 \three\beta_i v^i+{5 \over 2}\varepsilon U+
{5 \over 2}U^2+5\four\psi\Bigr\} \cr
&+{1 \over 2}\four h_{ij}U_{,ij}-{1 \over 8}
\Bigl(\three\tilde A_{ij} \three\tilde A_{ij}-{2 \over 3}\three K^2\Bigr).
 \cr}\eqno\eq
$$
The term relevant to the radiation reaction first appears in 
$\seven\psi$ and the equation for it becomes 
$$
\Delta_{flat} \seven\psi={1 \over 2}\five h_{ij}U_{,ij}.\eqno\eq
$$
Hence, $\seven\alpha$ may be also relevant to the radiation reaction 
and whether it may or not depends on the slice condition. 

From Eq.$\bbb$, the relation between 
$\three\tilde A_{ij}$ and $\three\beta_i$ becomes 
$$
-2\three\tilde A_{ij}+\three\beta_{i,j}
+\three\beta_{j,i}-{2 \over 3}\delta_{ij}\three\beta_{l,l}=0.\eqn\ppb
$$
$\three\tilde A_{ij}$ must also satisfy the momentum constraint. 
Since $\three\tilde A_{ij}$ does not contain the 
transverse-traceless(TT) part and only contains 
the longitudinal part, it can be written as 
$$
\three\tilde A_{ij} =\three W_{i,j}+\three W_{j,i}
-{2 \over 3}\delta_{ij}\three W_{k,k}~, \eqno\eq
$$
where $\three W_i$ is a vector on the 3D hypersurface and satisfies 
the momentum constraint at the first PN order as follows; 
$$
\Delta_{flat} \three W_i+{1 \over 3}\three W_{j,ji}
-{2 \over 3}\three K_{,i}=8\pi \rho v^i.
\eqn\momo
$$
From Eq.$\ppb$, the relation, 
$$
\three\beta_i=2\three W_i~,\eqno\eq
$$
holds and at the first PN order, Eq.$\kkg$ becomes 
$$
3\dot U=-\three K+\three\beta_{l,l}~,\eqn\dotu
$$
where $\dot U$ denotes the derivative of $U$ with respect to time, 
so that Eq.$\momo$ is rewritten as 
$$
\Delta_{flat} \three\beta_i=16\pi \rho v^i+\Bigl(\three K_{,i}-\dot U_{,i}
\Bigr).\eqn\betth
$$
This is the equation for the vector potential 
at the first PN order. 

From the next order, $\hbox{$_{(n)}$}\beta_i$ is determined by the gauge 
condition, $h_{ij,j}=0$. 
Making use of the momentum constraint and the 2PN order 
of Eq.$\kkg$, 
$$
6\four\dot\psi-3\three\beta_l U_{,l}-{1 \over 2}U(2\three K+3\dot U)
+\five K=\five\beta_{l,l}~, 
\eqno\eq
$$
the equation for $\five\beta_i$ is written as 
$$\eqalign{
\Delta_{flat} \five\beta_i &=16\pi\rho\Bigl[v^i \Bigl(v^2+2U+\varepsilon
+{P \over \rho}\Bigr)+\three\beta_i\Bigr]-8U_{,j}\three\tilde A_{ij} \cr
&+\five K_{,i}-U \three K_{,i}+{1 \over 3}U_{,i}\three K-2\four\dot\psi_{,i}
+{1 \over 2}(U \dot U)_{,i}+(\three\beta_l U_{,l})_{,i}~.\cr}
\eqn\betfi
$$
Since $J_i$ at the 1.5PN order vanishes, the merely geometrical equation 
for $\six\beta_i$ is given by 
$$
\Delta_{flat}\six\beta_i=\six K_{,i}~.
\eqn\betsi
$$

Then, let us consider the wave equation for $h_{ij}$. 
From Eqs.$\ggg$, $\kkk$, (2.21) and (2.27), the wave equation for $h_{ij}$ 
is written as 
$$\eqalign{
\square h_{ij}=&\Bigl(1-{\alpha^2 \over \psi^4}\Bigr)\Delta_{flat}h_{ij}
+\Bigl({\pa^2 \over \pa n^2}-{\pa^2 \over \pa t^2}\Bigr)h_{ij} \cr
&+{2\alpha \over \psi^4}
\Bigl[-{2\alpha \over \psi}\Bigl(\tilde D_i\tilde D_j
-{1 \over 3}\tilde\gamma_{ij}\tilde\Delta\Bigr)\psi
+{6\alpha \over \psi^2}\Bigl(\tilde D_i\psi \tilde D_j\psi
-{1 \over 3}\tilde\gamma_{ij}\tilde D_k\psi \tilde D^k\psi\Bigr) \cr
&~~~~~
-\Bigl(\tilde D_i\tilde D_j-{1 \over 3}\tilde\gamma_{ij}\tilde\Delta
\Bigr)\alpha
-{2 \over \psi}\Bigl(\tilde D_i\psi\tilde D_j\alpha
+\tilde D_j\psi\tilde D_i\alpha-{2 \over 3}\tilde\gamma_{ij}
\tilde D^k \psi\tilde D_k\alpha\Bigr)\Bigr] \cr
&+2\alpha^2 \Bigl(K \tilde A_{ij}-2\tilde A_{il} \tilde A^l_{~j}\Bigr)
+2\alpha \Bigl( \beta^m_{,~i} \tilde A_{mj}
+\beta^m_{~,j} \tilde A_{mi}-{2 \over 3} \beta^m_{~,m} \tilde A_{ij}
\Bigr)\cr
&-16\pi{\alpha^2 \over \psi^4}
\Bigl(S_{ij}-{1 \over 3}\gamma_{ij}S^l_{~l}\Bigr)
-{\pa \over \pa n}\Bigl( \beta^m_{~,i} \tilde \gamma_{mj}
+ \beta^m_{~,j} \tilde \gamma_{mi}
-{2 \over 3}\beta^m_{~,m} \tilde \gamma_{ij} \Bigr)
+2{\pa\alpha \over \pa n}\tilde A_{ij} \cr 
\equiv&\tau_{ij}, \cr}
\eqn\waveeq
$$
where 
$$
\square =-{\pa^2 \over \pa t^2}+\Delta_{flat}.\eqno\eq
$$
We should note that $\four\tau_{ij}$ has the TT property, \ie, 
$\four\tau_{ij,j}=0$ and $\four\tau_{ii}=0$. 
This is a natural consequence of the transverse gauge, $h_{ij,j}=0$ and 
$h_{ii}=O(h^2)$. 
Thus $\four h_{ij}$ is determined\footnote{*}{Since $O(h^2)$ turns out 
to be $O(c^{-8})$, it is enough to consider only the linear order of 
$h_{ij}$ in the case when we perform the PN approximation 
up to the 3.5PN order.} from 
$$
\Delta_{flat} \four h_{ij}=\four\tau_{ij}. 
\eqn\fourh
$$
$\five h_{ij}$ is derived from 
$$
\five h_{ij}(t)={1 \over 4\pi}
{\pa \over \pa t}\int \four\tau_{ij}(t, {\bf y})d^3y, 
\eqn\quadr
$$
and the quadrupole mode of gravitational waves in the wave zone 
is written as 
$$
h_{ij}^{rad}=-{1 \over 4\pi} \lim_{|{\bf x}| \rightarrow \infty}
\int {\four\tau_{ij}(t-|{\bf x}-{\bf y}|, {\bf y}) \over 
|{\bf x}-{\bf y}|}d^3y.
\eqn\quadra
$$
In the subsequent section, we derive the quadrupole radiation-reaction 
metric in the near zone using Eq.$\quadr$. 

Finally, we show the evolution equation for $K$. 
Since we adopt slice conditions which do not satisfy $K=0$(\ie, maximal slice 
condition), the evolution equation for $K$ is necessary. 
The evolution equations appear at the 1PN, 2PN and 
2.5PN orders which become respectively  
$$
{\pa \over \pa t}\three K=4 \pi\rho\Bigl(2v^2+\varepsilon+2U+3 {P \over \rho}
\Bigr)-\Delta_{flat}X, {\hskip 5.5cm}
\eqn\Kth
$$
$$
\eqalign{
{\pa \over \pa t}\five K=&4 \pi\rho\Bigl[2v^4+v^2\Bigl(6U+2\varepsilon
+2{P \over \rho}\Bigr)-\Bigl(\varepsilon +{3P \over \rho}\Bigr)U 
-4 U^2+4\four\psi+X+4\three\beta_i v^i \Bigr] \cr
&+\three \tilde A_{ij}\three \tilde A_{ij}+{1 \over 3}\three K^2
-\four h_{ij}U_{,ij}
+\three\beta_i \three K_{,i} \cr
&-{3 \over 2}UU_{,k} U_{,k}-U_{,k}X_{,k}+2U_{,k}\four\psi_{,k}
-\Delta_{flat}\six\alpha+2U\Delta_{flat}X, \cr}
\eqno\eq
$$
$$
{\pa \over \pa t}\six K=-\Delta_{flat}\seven\alpha-\five h_{ij}U_{,ij}. 
{\hskip 8cm}
\eqn\dtsixK
$$

We note that for the PN equations of motion up to the 2.5PN order, 
we need $\two\alpha$, $\four\alpha$, $\six\alpha$, $\seven\alpha$, 
$\two\psi$, $\four\psi$, $\three\beta_i$, $\five\beta_i$, $\six\beta_i$, 
$\four h_{ij}$, $\five h_{ij}$, $\three K$, $\five K$ and $\six K$. 
Therefore, if we solve the above set of the equations, we can obtain 
the 2.5 PN equations of motion. 
Up to the 2.5PN order, the hydrodynamic equations become 
$$
\eqalignno{
{\pa S_i \over \pa t}+{\pa (S_i v^j) \over \pa x^j}
&=-\Bigl(1+2 U+{5 \over 4}U^2+6\four\psi+X\Bigr)P_{,i} \cr
&+\rho_{\ast}\Bigl[U_{,i} \Bigl\{1+\varepsilon+{P \over \rho}
+{3 \over 2}v^2-U+{5 \over 8}v^4+4 v^2U
+\Bigl({3 \over 2}v^2-U\Bigr)\Bigl(\varepsilon+{P \over \rho}\Bigr)
+3\three\beta_j v^j \Bigr\} \cr
&-X_{,i}\Bigl(1+\varepsilon+{P \over \rho}+{v^2 \over 2}\Bigr) 
+2v^2\four\psi_{,i}-\six\alpha_{,i}-\seven\alpha_{,i} \cr
&+v^j \Bigl\{ \three\beta_{j,i}\Bigl(1+\varepsilon+{P \over \rho}
+{v^2 \over 2}+3U\Bigr)+\five\beta_{j,i}+\six\beta_{j,i} \Bigr\}
+\three\beta_j\three\beta_{j,i} \cr
&+{1 \over 2}v^jv^k(\four h_{jk,i}+\five h_{jk,i}) +O(c^{-8}) 
\Bigr], &\eq \cr
{\pa H \over \pa t}+{\pa (H v^j) \over \pa x^j} 
&=-P\Bigl[v^j_{~,j}+{\pa \over \pa t}\Bigl({1 \over 2}v^2+3 U\Bigr)
+{\pa \over \pa x^j} \Bigl\{\Bigl({1 \over 2}v^2+3 U\Bigr)v^j \Bigr\} 
+O(c^{-5}) \Bigr], &\eq \cr}
$$
where we make use of relations 
$$\eqalign{
&\alpha S^0=\rho_* \Bigl[1+\varepsilon+{P \over \rho}+{v^2 \over 2}
+{v^2 \over 2}\Bigl(\varepsilon+{P \over \rho}\Bigr)+{3 \over 8}v^4
+2v^2U+\three\beta_j v^j ++O(c^{-6})\Bigr],\cr
&S_i=\rho_*\Bigl[v^i 
\Bigl(1+\varepsilon+{P \over \rho}+{v^2 \over 2}+3U\Bigr)+\three \beta_i
+O(c^{-5})\Bigr].\cr}\eqno\eq
$$

\chapter{Slice Conditions}

In this section, we perform the PN analysis using 
the conformal slice\refmark\conf, maximal slice and 
harmonic slice\refmark\masso which are often used 
in 3D numerical relativity. 
Among them, we find that the conformal slice seems most tractable 
and useful to estimate gravitational waves in the far zone, 
while the maximal slice is suitable for describing 
the equilibrium configurations. 
Hence, first of all we describe the property of the conformal slice 
and then mention the properties of other slices.

\section{Conformal Slice}

The conformal slice is defined\refmark\conf as 
$$
\alpha=\exp\Bigl(-2\epsilon-{2 \over 3}\epsilon^3-{2 \over 5}\epsilon^5
\Bigr),\eqno\eq
$$
where $\epsilon=\psi-1$. 
In the conformal slice, $\hbox{$_{(n)}$}\alpha$ becomes 
$$\eqalign{
&\two\alpha=-2\two\psi,\cr
&\four\alpha=2(\two\psi)^2-2\four\psi,\cr
&\six\alpha=-2(\two\psi)^3+4\two\psi \four\psi-2\six\psi,\cr
&\seven\alpha=-2\seven\psi,\cr}\eqno\eq
$$
Although in the usual PN approximation we need to solve 
the Poisson equation for the lapse function, this slicing saves 
solving it. 

In the conformal slice, equations $\kkk$ and $\kkt$ are rewritten as 
$$\eqalign{
{\pa \over \pa n} \tilde A_{ij}=&-{1 \over 2}{\alpha \over \psi^4}
\Delta_{flat} h_{ij}
+{2\alpha \over \psi^4}
\Bigl[ \Bigl(\tilde D_i \tilde D_j \psi-
 {\tilde \gamma_{ij} \over 3}\tilde D_k \tilde D^k \psi \Bigr)
{\epsilon \over \psi} (1+\epsilon+\epsilon^2+\epsilon^3+\epsilon^4) \cr
&~~~~~-{1 \over \psi^2}\Bigl(\tilde D_i \psi\tilde D_j\psi
-{1 \over 3}\tilde \gamma_{ij}\tilde D^k \psi \tilde D_k \psi\Bigr)\cr
&~~~~~~~~~~~~~~(3+6\epsilon+6\epsilon^2+6\epsilon^3+6\epsilon^4+12\epsilon^5
+10\epsilon^6+8\epsilon^7+6\epsilon^8+4\epsilon^9+2\epsilon^{10})
 \Bigr]\cr
&+\alpha (K \tilde A_{ij}-2\tilde A_{il} \tilde A^l_{~j}) 
+ \beta^m_{~,i} \tilde A_{mj}+ \beta^m_{~,j} \tilde A_{mi}
-{2 \over 3} \beta^m_{~,m} \tilde A_{ij} \cr
&-8\pi{\alpha \over \psi^4}
\Bigl(S_{ij}-{1 \over 3}\gamma_{ij}S^l_{~l}\Bigr),\cr}
 \eqn\kkkq
$$
and 
$$\eqalign{
{\pa K \over \pa n}=&2 {\alpha \over \psi^4}\Bigl[\tilde \Delta \psi 
(1+\epsilon^2+\epsilon^4)-{2 \over \psi^2}\tilde D_k \psi \tilde D^k \psi 
(3\epsilon^5+2\epsilon^6+2\epsilon^7+\epsilon^8+\epsilon^9)
\Bigr]\cr
&+\alpha\Bigl(\tilde A_{ij} \tilde A^{ij}+{1 \over 3}K^2 \Bigr)
+4\pi \alpha(S^l_{~l}+\rho_{H}),\cr}\eqno\eq
$$
where we use the TT property as well as the linear approximation 
for $h_{ij}$ in the above equation. 
Then Eq.$\waveeq$ is written as
$$\eqalign{
\square h_{ij}=&-\Bigl({\alpha^2 \over \psi^4}-1\Bigr)\Delta_{flat} h_{ij}
+\Bigl({\pa^2 \over \pa n^2}-{\pa^2 \over \pa t^2}\Bigr)
h_{ij}\cr
&+{4\alpha^2 \over \psi^4}
\Bigl[ \Bigl(\tilde D_i \tilde D_j \psi-
 {\tilde \gamma_{ij} \over 3}\tilde D_k \tilde D^k \psi \Bigr)
{\epsilon \over \psi} (1+\epsilon+\epsilon^2+\epsilon^3+\epsilon^4) \cr
&~~~~-{1 \over \psi^2}\Bigl(\tilde D_i \psi\tilde D_j\psi
-{1 \over 3}\tilde \gamma_{ij}\tilde D^k \psi \tilde D_k \psi\Bigr)\cr
&~~~~~~~~(3+6\epsilon+6\epsilon^2+6\epsilon^3+6\epsilon^4+12\epsilon^5
+10\epsilon^6+8\epsilon^7+6\epsilon^8+4\epsilon^9+2\epsilon^{10})
\Bigr] \cr
&+2\alpha^2 (K \tilde A_{ij}-2\tilde A_{il} \tilde A^l_{~j})
+2\alpha \Bigl( \beta^m_{~,i} \tilde A_{mj}
+\beta^m_{~,j} \tilde A_{mi}-{2 \over 3} \beta^m_{~,m} \tilde A_{ij}
\Bigr)\cr
&-16\pi{\alpha^2 \over \psi^4}
\Bigl(S_{ij}-{1 \over 3}\gamma_{ij}S^l_{~l}\Bigr)
-{d \over dt}\Bigl( \beta^m_{~,i} \tilde \gamma_{mj}
+ \beta^m_{~,j} \tilde \gamma_{mi}
-{2 \over 3}\beta^m_{~,m} \tilde \gamma_{ij} \Bigr)
+2{\pa\alpha \over \pa n}\tilde A_{ij} \cr 
\equiv &\tau_{ij},\cr} \eqn\waveq
$$
where  we use $\epsilon=\psi-1$ and $\psi$ satisfies 
$$
\tilde \Delta \psi=-2\pi \rho_{H}\psi^5-{\psi^5 \over 8}
\Bigl(\tilde A_{ij} \tilde A^{ij}-{2 \over 3}K^2 \Bigr).\eqn\confa
$$
Eq.$\waveq$ is expanded as follows; 
$$
\eqalign{
\square h_{ij}
=&\Bigl(U U_{,ij}-{1 \over 3}\delta_{ij}U \Delta_{flat}U
-3U_{,i} U_{,j}+\delta_{ij}U_{,k} U_{,k} \Bigr) \cr
&-16\pi \Bigl(\rho v^i v^j -{1 \over 3}\delta_{ij}\rho v^2 \Bigr)
-\Bigl(\three\dot\beta_{i,j}+\three\dot\beta_{j,i}
-{2 \over 3}\delta_{ij}\three\dot\beta_{k,k} \Bigr)+O(c^{-6}) \cr
=&\four\tau_{ij}+O(c^{-6}),\cr} 
\eqn\qua
$$
where we use $\epsilon=2U$ which holds in the PN approximation. 

In the conformal slice, the asymptotic form of $\epsilon$ becomes 
$$
\epsilon\sim {M_{ADM} \over 2r}. 
\eqno\eq
$$
Thus $\alpha$ behaves as $1-M_{ADM}/r$ at spatial infinity. 
This means that in the conformal slice, the metric at spatial infinity 
becomes the static Schwarzschild's one. 
This property seems helpful for discerning the wave part from 
the non-wave part in the wave zone in numerical relativity. 

Also, we have an advantage to derive the radiation reaction potential 
in this slice; 
From a relation $\seven\alpha=-2\seven\psi$ and Eq.(2.45), we have 
$$
\Delta_{flat} \seven\alpha=-\five h_{ij}(t)U_{,ij}.\eqno\eq
$$
Thus, the radiation reaction potential $\seven \alpha$ is derived as 
$$
\seven \alpha ={\five h_{ij}(t) \over 4\pi}\int U_{,ij} d^3x.\eqno\eq
$$


Finally, we comment on the following week point of the conformal slice; 
in the conformal slice, the evolution equation for $\three K$ becomes 
$$
\three\dot K=4\pi\rho\Bigl(v^2+3{P \over \rho}-{1 \over 2}U\Bigr).  
\eqno\eq
$$
Since $\dot K$ does not vanish, $K$ continues to change 
even in the case of a stationary spacetime. 
Thus, it seems inconvenient to describe equilibrium configurations of stars 
and binary systems in the conformal slice. 
To describe equiriblium configurations, we had better use the slice, 
such as the maximal slice, where $\dot K=0$ is satisfied.

\section{Maximal Slice}

The maximal slice is given by
$$
K=0 ,
\eqno\eq
$$
and this equation leads to the equation for $\alpha$ as
$$
D_k D^k\alpha=\alpha\Bigl(\tilde A_{ij} \tilde A^{ij}+4\pi (E+S^l_{~l})
\Bigr).\eqn\aap
$$
At the first PN order, the equation becomes 
$$
\Delta_{flat}\four X_{MS}=4\pi\rho \Bigl(2 v^2+\varepsilon+{3P \over \rho}
+2U \Bigr), 
\eqno\eq
$$
where the subscript $MS$ denotes ``maximal slice''. 
In the case of the conformal slice, the following relation holds; 
$$
X_{CS}=-2\four\psi,\eqno\eq
$$
where $CS$ similarly denotes ``conformal slice''. 
Using the above equation, we rewrite $X_{MS}$ as 
$$
X_{MS}=-2\four\psi+Y. 
\eqno\eq
$$
Then the equation for $Y$ becomes
$$
\Delta_{flat} Y=4\pi\Bigl(\rho v^2+3P-{1 \over 2}\rho U\Bigr).\eqno\eq
$$

We should also note that by means of the virial theorem\refmark\ellp, 
the integration of the source term for $Y$ can be written as 
$$
\int \Bigl(\rho v^2+3P-{1 \over 2}\rho U\Bigr)d^3x 
={1 \over 2}\ddot I_{ll},\eqno\eq
$$
where
$$
I_{ij}(t)=\int \rho x^i x^j d^3x.\eqno\eq
$$
Hence, the behavior of $Y$ far from the matter becomes 
$$
Y \sim -{1 \over 2r} \ddot I_{ll}.\eqno\eq
$$
In total, the behavior of $\alpha$ in the wave zone becomes 
$$
\alpha \sim 1-{1 \over r}\Bigl(M+{1 \over 2}\ddot I_{ll}\Bigr).
\eqno\eq
$$
Therefore, contrary to the conformal slice, 
in the maximal slice, the spurious time-dependent term 
is included in $\alpha$ in the wave zone. 
Since the metric does not approach the static Schwarzschild metric 
even in spatial infinity, 
the maximal slice is inconvenient to distinguish a wave part from 
non-wave parts such as the Newtonian potential.

In the case of the maximal slice, the equations for 
the shift vector are obtained by simply taking $K=0$ in Eqs.$\betth$ 
and $\betfi$.
Also, it is found that the equation for $\seven \alpha$ is the same as that 
in the conformal slice: The right-hand side of Eq.$\aap$ 
has no $O(c^{-7})$ terms. Therefore Eq.$\aap$ becomes 
$$
\Delta_{flat} \seven\alpha=-\five h_{ij} U_{,ij} .
\eqno\eq
$$

Finally, we show the wave equation for $h_{ij}$ 
in the maximal slice as 
$$
\eqalign{
\square h_{ij}
=&-2\Bigl(Y_{,ij}-{1 \over 3}\delta_{ij}\Delta_{flat} Y\Bigr)
+\Bigl(U U_{,ij}-{1 \over 3}\delta_{ij}U \Delta_{flat}U
-3U_{,i} U_{,j}+\delta_{ij}U_{,k} U_{,k} \Bigr) \cr
&-16\pi \Bigl(\rho v^i v^j -{1 \over 3}\delta_{ij}\rho v^2 \Bigr)
-\Bigl(\three\dot\beta_{i,j}+\three\dot\beta_{j,i}
-{2 \over 3}\delta_{ij}\three\dot\beta_{k,k} \Bigr)+O(c^{-6}). \cr}
\eqn\qua
$$

\section{Harmonic Slice}

The condition for the harmonic slice is 
$$
\square t=0, 
\eqno\eq
$$
which becomes in the (3+1) terminology, 
$$
\dot\alpha+\alpha^2 K-\beta^i \alpha_{,i}=0. 
\eqno\eq
$$
Differentiating this equation with respect to time, 
the wave equation for the lapse function is derived as 
$$
\eqalign{
\square\alpha=
&4\pi\alpha^3(S^l_{~l}-3\rho_{H})-\Bigl({\alpha^2 \over \psi^4}\tilde\Delta
-\Delta_{flat}\Bigr)
\alpha-{2\alpha^2 \over \psi^5}\tilde D^l \psi \tilde D_l \alpha
-{8 \alpha^3 \over \psi^5}\tilde\Delta\psi \cr
&+2\alpha\dot\alpha K+{\alpha^3 \over \psi^4}\tilde R+\alpha^3 K^2 
+\alpha^2\beta^i\tilde D_i K-\dot\beta^i\alpha_{,i} 
-\beta^i \dot\alpha_{,i} \cr
\equiv &\Lambda_{\alpha}, \cr}
\eqno\eq
$$
where $\Lambda_{\alpha}$ is expanded as follows,
$$
\Lambda_{\alpha}=4\pi\rho \Bigl[1+\Bigl(v^2+3{P \over \rho}-{U \over 2}
\Bigr)\Bigr]+\Delta_{flat} \Bigl({U^2 \over 2}-2\four\psi\Bigr)+O(c^{-6}). 
\eqno\eq
$$
This equation is formally solved by using the retarded Green 
function and the Taylor expansion.
For example, we obtain the Newtonian and first PN order lapse function
$$
\eqalignno{
\two\alpha=&-\int d^3y {\rho(t,{\bf y}) \over 
\vert {\bf x}-{\bf y} \vert}=-{1 \over r}\int\rho d^3y+O(r^{-2}), &\eq \cr
\four\alpha=&-{1 \over 2}\int d^3y \ddot \rho \vert {\bf x}-{\bf y} \vert
-\int d^3y {(\rho v^2+3P-{1 \over 2}\rho U) \over 
\vert {\bf x}-{\bf y} \vert} +{1 \over 2}U^2-2 \four\psi. &\eq \cr}
$$
Thus, at the spatial infinity, we find the following behavior
$$
\four\alpha+2\four\psi\sim -{3 \over 4r}\Bigl(\ddot I_{kk}
-{1 \over 3}n^k n^l \ddot I_{kl}\Bigr) , 
\eqno\eq
$$
where $n^i=x^i/r$.  
From these equations we find that at the spatial infinity 
the lapse function does not behave as $1-M/r+O(r^{-2})$ unlike 
in the conformal slice, but behaves as $1-({M+T(t)})/r+O(r^{-2})$.
Thus the harmonic slice is also inconvenient to distinguish a wave part 
from non-wave parts. 

The quadrupole radiation reaction potential takes the following 
rather lengthy form. 
$$
\eqalign{
\seven\alpha=&{1 \over 480\pi}{\pa^5 \over \pa t^5}\int d^3y \rho
\vert {\bf x}-{\bf y} \vert^4
+{1 \over 24\pi}{\pa^3 \over \pa t^3}\int d^3y\four\Lambda_{\alpha}
\vert {\bf x}-{\bf y} \vert^2+{1 \over 4\pi}{\pa \over \pa t}\int d^3y
\six\Lambda_{\alpha} \cr
&+{1 \over 4\pi}\int d^3y{\five h_{ij} U_{,ij} \over 
\vert {\bf x}-{\bf y} \vert}. \cr}
\eqno\eq
$$
This expression is similar to Chandrasekhar's 
one in the harmonic gauge\refmark\chandra and indicates that the fifth 
time derivative of the quadrupole moment appears in the 
reaction force, which is not convenient 
to treat in numerical calculations.

\chapter{The Radiation Reaction due to Quadrupole Radiation} 


This topic has been already investigated by using some gauge conditions 
in previous papers\refmark{\chandra,\schafer,\bds}. 
However, if we use the combination of the conformal slice and the 
transverse gauge, calculations are simplified. This is why we briefly mention 
the derivation of the radiation reaction potential in this section. 

In combination of the conformal slice and the transverse gauge, 
Eq.$\quadr$ becomes
$$
\eqalign{\five h_{ij}(t)=&{1 \over 4\pi}
{\pa \over \pa t}\int \Bigl[ -16\pi \Bigl
(\rho v^i v^j -{1 \over 3}\delta_{ij}\rho v^2 \Bigr) \cr
&~~~~~~~~+\Bigl(U U_{,ij}-{1 \over 3}\delta_{ij}U \Delta_{flat}U
-3U_{,i} U_{,j}+\delta_{ij}U_{,k} U_{,k} \Bigr) \Bigr] d^3y \cr
&+{1 \over 4 \pi}{\pa \over \pa t}\int\Bigl(\three\dot\beta_{i,j}
+\three\dot\beta_{j,i}-{2 \over 3}\delta_{ij}\three\dot\beta_{k,k} \Bigr)
d^3y .}
\eqn\quadrup
$$
From a straightforward calculation, we find that the sum of 
the first and second lines becomes $-2\bI_{ij}^{(3)}$ and 
the third line becomes $6\bI_{ij}^{(3)} /5$, where 
$\bI_{ij}^{(3)}=d^3 \bI_{ij}/dt^3$.  
(This calculation is replaced by a fairly simple one 
noticing the transverse property of $\four\tau_{ij}$. 
It is described in the appendix A.)
Thus, $\five h_{ij}$ in the near zone becomes 
$$
\five h_{ij}=-{4 \over 5}\bI_{ij}^{(3)} , \eqn\nearh
$$
where 
$$
\bI_{ij}=I_{ij}-{1 \over 3}\delta_{ij} I_{kk}.\eqno\eq
$$
Since $h_{ij}$ has the transverse and traceless property, 
it is likely that $\five h_{ij}$ 
remains the same for other slices.
However it is not clear whether the TT property of $h_{ij}$ is satisfied 
even after the PN expansion is taken in the near zone and, as a result, 
whether $\five h_{ij}$ is independent of slicing conditions or not. 
The fact that slicing conditions never affect $\five h_{ij}$ is understood 
on the ground that $\four\tau_{ij}$ does not depend on slices, 
which will be shown in the section 5. 

Then the Hamiltonian constraint at the 2.5PN order turns out to be
$$
\Delta_{flat} \seven\psi=-{2 \over 5} \bI_{ij}^{(3)} U_{,ij}
={1 \over 5} \bI_{ij}^{(3)} \Delta_{flat}\chi_{,ij} \; , \eqno\eq
$$
where $\chi$ is the superpotential\refmark\chand and defined as 
$$
\chi=-\int\rho\vert {\bf x}-{\bf y} \vert d^3y, 
\eqno\eq
$$
which satisfies the relation $\Delta_{flat}\chi=-2U$. 
From this, we find $\seven\psi$ takes the following form,
$$\eqalign{
\seven\psi&=-{1 \over 5} \bI_{ij}^{(3)} \int\rho_{,i} 
{(x^j-y^j)\over \vert {\bf x}-{\bf y} \vert} d^3y\cr
&={1 \over 5} \bI_{ij}^{(3)}\biggl(-x^j U_{,i}
+\int{\rho_{,i}y^j \over \vert {\bf x}-{\bf y} \vert} d^3y\biggr).\cr} 
\eqn\psisev
$$
Therefore, the lapse function at the 2.5PN order, 
$\seven\alpha=-2\seven\psi$, is derived from $U$ and $U_r$, where 
$U_r$ satisfies\refmark\bds
$$
\Delta U_r=-4\pi \bI_{ij}^{(3)}\rho_{,i}x^j. \eqn\alpsev
$$
Since the right-hand side of Eq.$\dtsixK$ cancels out, $\six K$ disappears 
if the $\six K$ does not exist on the initial hypersurface, 
which seems reasonable under the condition that there are 
no initial gravitational waves. 
Also, $\six\beta_i$ vanishes according to Eq.$\betsi$. 
Hence, the quadrupole 
radiation reaction metric has the same form as that derived in the case of the 
maximal slice\refmark{\schafer,\bds}. 

From Eq.$\EOM$, the PN equations of motion becomes 
$$
\dot v^i+v^j v^i_{,~j}=-{P_{,i} \over \rho}+U_{,i}+F_i^{1PN}
+F_i^{2PN}+F_i^{2.5PN}+O(c^{-8}) ,
\eqno\eq
$$
where $F_i^{1PN}$ and $F_i^{2PN}$ are, respectively, the 1PN and 2PN forces 
and conservative ones.  
Since the radiation reaction potentials, $\five h_{ij}$ and $\seven\alpha$, 
are the same as those by Sch\"afer\refmark\schafer and 
Blanchet, Damour and Sch\"afer\refmark\bds in which they use the ADM gauge, 
the radiation reaction force per unit mass, 
$F_i^{2.5PN} \equiv F_i^{r}$, is the same as their force and 
$$
\eqalign{F_i^{r}&=-\Bigl( (\five h_{ij} v^{j})^{\cdot}+
v^k v^j_{,k} \five h_{ij}+\seven\alpha_{,i} \Bigr) \cr
&=\Bigl[ {4\over5}(\bI_{ij}^{(3)} v^j)^{\cdot}+{4\over5} \bI_{ij}^{(3)} 
v^k v^j_{,k}+{2\over5}\bI_{kl}^{(3)} {\pa \over \pa x^i}
\int\rho(t,{\bf y}) { (x^k-y^k)(x^l-y^l) \over \vert 
{\bf x}-{\bf y} \vert^3} d^3y \Bigr]. \cr}
\eqn\rrforce
$$
Since the work done by the force $\rrforce$ is given by 
$$
\eqalign{W &\equiv\int \rho F_i^{r} v^i d^3x \cr
&={4\over5}{\pa \over \pa t}\Bigl(\bI_{ij}^{(3)} \int \rho v^i v^j 
d^3x \Bigr) -{1\over5}\bI_{ij}^{(3)}\bI^{(3) ij}, } 
\eqn\dEquadr
$$
we obtain the so-called quadrupole formula of the energy loss 
by averaging Eq.$\dEquadr$ 
with respect to time as 
$$
\left\langle {dE_N \over dt} \right\rangle=-{1\over5}
\left\langle \bI_{ij}^{(3)} \bI^{(3) ij} \right\rangle+O(c^{-6}). 
\eqno\eq
$$

\chapter{Strategy to obtain 2PN tensor potential}

In this section, we describe methods to solve 
the equation for the 2PN tensor potential $\four h_{ij}$.
Although Eq.$\fourh$ is formally solved as 
$$
\four h_{ij}(t,{\bf x})=-{1 \over 4\pi}\int{\four\tau_{ij}(t,{\bf y}) 
\over \vert {\bf x}-{\bf y} \vert}d^3y, 
\eqno\eq
$$
it seems difficult to estimate this integral in practice since
$\four \tau_{ij} \rightarrow O(r^{-3})$ for $r \rightarrow \infty$ and the
integral is taken all over the space.
Thus it is desirable to replace this equation by some tractable forms 
in numerical evaluation. 
In the following, we show two approaches: 
In section 5.1, we change Eq.(5.1) into the form in which 
the integration is performed only over the matter distribution like as
in the Newtonian potential. In section 5.2, we propose a method to solve 
Eq.$\fourh$ as the boundary value problem. 

\section{Direct integration method}

The explicit form of $\four \tau_{ij}$ is 
$$
\eqalign{
\four\tau_{ij}=&-2\hat\pa_{ij}(X+2\four\psi)+U\hat\pa_{ij} U 
-3U_{,i}U_{,j}+\delta_{ij}U_{,k}U_{,k}
-16\pi\Bigl(\rho v^iv^j-{1 \over 3}\delta_{ij}\rho v^2\Bigr) \cr
&-\Bigl( \three\dot\beta_{i,j}+\three\dot\beta_{j,i}
-{2 \over 3}\delta_{ij}\three\dot\beta_{k,k} \Bigr), \cr}
\eqno\eq
$$
where
$$
\hat\pa_{ij} \equiv {\pa^2 \over \pa x^i \pa x^j}
-{1 \over 3}\delta_{ij}\Delta_{flat}.
$$ 
Although $\four\tau_{ij}$ looks as if it depends on the slice condition, 
the independence is shown as follows.
Eq.$\betth$ is rewritten as 
$$
\Delta_{flat}\three\beta_i=\Delta_{flat}p_i+\three K_{,i},  
\eqno\eq
$$
where 
$$
p_i=-4\int{\rho v^i \over \vert {\bf x}-{\bf y} \vert} d^3y
-{1 \over 2}\Bigl(\int \dot\rho \vert {\bf x}-{\bf y} \vert d^3y 
\Bigr)_{,i}~. 
\eqno\eq
$$
This is solved as
$$
\three\beta_i=p_i-{1 \over 4\pi}\Bigl( \int {\three K \over 
\vert {\bf x}-{\bf y} \vert} d^3y \Bigr)_{,i}~. 
\eqn\thbet 
$$
From Eqs.$\psifo$ and $\Kth$, we obtain 
$$
\three\dot K=-\Delta_{flat}(X+2\four\psi)+4\pi\rho\Bigl( v^2
+3{P \over \rho}-{U \over 2} \Bigr). 
\eqn\abceq
$$
Combining Eq.$\thbet$ with Eq.$\abceq$, the equation for 
$\three\dot\beta_i$ is written as  
$$
\three\dot\beta_i=\dot  p_i-(X+2\four\psi)_{,i}
-\Bigl\{\int{\Bigl( \rho v^2+3P-\rho U / 2\Bigr) 
\over \vert {\bf x}-{\bf y} \vert}d^3y \Bigr\}_{,i} ~.
\eqno\eq
$$
Using this relation, 
the source term, $\four\tau_{ij}$, is split into five parts 
$$
\four\tau_{ij}=
\four\tau_{ij}^{(S)}+\four\tau_{ij}^{(U)}+\four\tau_{ij}^{(C)}
+\four\tau_{ij}^{(\rho)}+\four\tau_{ij}^{(V)}, 
\eqno\eq
$$
where we introduce 
$$
\eqalign{
\four\tau_{ij}^{(S)}=&-16\pi\Bigl( \rho v^iv^j-{1 \over 3}\delta_{ij}
\rho v^2 \Bigr), \cr
\four\tau_{ij}^{(U)}=&UU_{,ij}-{1 \over 3}\delta_{ij}U\Delta_{flat} U
-3U_{,i}U_{,j}+\delta_{ij}U_{,k}U_{,k}, \cr
\four\tau_{ij}^{(C)}=&4{\pa \over \pa x^j}\int{(\rho v^i)^{\cdot}
\over \vert {\bf x}-{\bf y} \vert}d^3y
+4{\pa \over \pa x^i}\int{(\rho v^j)^{\cdot}
\over \vert {\bf x}-{\bf y} \vert}d^3y
-{8 \over 3}\delta_{ij}{\pa \over \pa x^k}\int{(\rho v^k)^{\cdot}
\over \vert {\bf x}-{\bf y} \vert}d^3y, \cr
\four\tau_{ij}^{(\rho)}=&\hat\pa_{ij}\int\ddot\rho \vert {\bf x}-{\bf y} 
\vert d^3y, \cr
\four\tau_{ij}^{(V)}=&2\hat\pa_{ij}\int{\Bigl( \rho v^2+3P
-\rho U / 2\Bigr) \over \vert {\bf x}-{\bf y} \vert}d^3y. \cr} 
\eqno\eq
$$
Thus it becomes clear that $\four h_{ij}$ and $\five h_{ij}$ 
as well as $\four\tau_{ij}$ are expressed in terms of matter variables 
only and thus do not depend on slicing conditions. 

Then, we define 
$\Delta_{flat}\four h_{ij}^{(S)}=\four\tau_{ij}^{(S)}$, 
$\Delta_{flat}\four h_{ij}^{(U)}=\four\tau_{ij}^{(U)}$, 
$\Delta_{flat}\four h_{ij}^{(C)}=\four\tau_{ij}^{(C)}$, 
$\Delta_{flat}\four h_{ij}^{(\rho)}=\four\tau_{ij}^{(\rho)}$ and 
$\Delta_{flat}\four h_{ij}^{(V)}=\four\tau_{ij}^{(V)}$, 
and consider each term separately.
First, since $\four\tau_{ij}^{(S)}$ is a compact source, 
we immediately obtain  
$$
\four h_{ij}^{(S)}=4\int{\Bigl(\rho v^i v^j-{1 \over 3}\delta_{ij}
\rho v^2 \Bigr) \over \vert {\bf x}-{\bf y} \vert}d^3y. 
$$
Second, we consider the following equation 
$$
\Delta_{flat} G({\bf x},{\bf y_1},{\bf y_2})={1 \over \vert {\bf x}-{\bf y_1} 
\vert \vert {\bf x}-{\bf y_2} \vert}. 
\eqn\G
$$
It is possible to write $\four h_{ij}^{(U)}$ using integrals over the matter 
if this function, $G$, is used.
Eq.$\G$ has solutions\refmark\ohta,
$$
G({\bf x},{\bf y_1},{\bf y_2})=\ln(r_1+r_2 \pm r_{12}), 
\eqno\eq
$$
where
$$
\eqalign{
r_1=&\vert {\bf x}-{\bf y_1} \vert, \cr
r_2=&\vert {\bf x}-{\bf y_2} \vert, \cr
r_{12}=&\vert {\bf y_1}-{\bf y_2} \vert. \cr}
\eqno\eq
$$
Note that $\ln(r_1+r_2 - r_{12})$ is not regular on 
the interval between ${\bf y_1}$ and ${\bf y_2}$, 
while $\ln(r_1+r_2+r_{12})$ is regular on the matter. 
Thus we use $\ln (r_1+r_2+r_{12})$ as a Green function. 
Using this function, $UU_{,ij}$ and $U_{,i}U_{,j}$ are rewritten as 
$$
\eqalign{
UU_{,ij}=&
\Bigl[ {\pa^2 \over \pa x^i \pa x^j}\Bigl(\int{\rho({\bf y_1}) \over 
|{\bf x}-{\bf y_1}|} d^3y_1 \Bigr) \Bigr] 
\Bigl(\int{\rho({\bf y_2}) \over |{\bf x}-{\bf y_2}|} d^3y_2 
\Bigr) \cr
=&\int d^3y_1 d^3y_2 \rho({\bf y_1})\rho({\bf y_2})
{\pa ^2 \over \pa y_1^i \pa y_1^j}\Bigl({1 \over |{\bf x}-{\bf y_1}|
|{\bf x}-{\bf y_2}|} \Bigr) \cr
=&\Delta_{flat} \int d^3y_1 d^3y_2 \rho({\bf y_1}) \rho({\bf y_2}) 
{\pa^2 \over \pa y_1^i \pa y_1^j} \ln(r_1+r_2+r_{12}), \cr
U_{,i}U_{,j}=&
\Bigl({\pa \over \pa x^i}\int{\rho({\bf y_1}) \over 
|{\bf x}-{\bf y_1}|} d^3y_1 \Bigr) 
\Bigl({\pa \over \pa x^j}\int{\rho({\bf y_2}) \over 
|{\bf x}-{\bf y_2}|} d^3y_2 \Bigr) \cr
=&\int d^3y_1 d^3y_2 \rho({\bf y_1})\rho({\bf y_2})
{\pa ^2 \over \pa y_1^i \pa y_2^j}\Bigl({1 \over |{\bf x}-{\bf y_1}|
|{\bf x}-{\bf y_2}|} \Bigr) \cr
=&\Delta_{flat} \int d^3y_1 d^3y_2 \rho({\bf y_1}) \rho({\bf y_2}) 
{\pa^2 \over \pa y_1^i \pa y_2^j} \ln(r_1+r_2+r_{12}). \cr}
\eqno\eq
$$
Thus we can express $\four h_{ij}^{(U)}$ using the integral 
over the matter as 
$$
\eqalign{
\four h_{ij}^{(U)}=\int d^3y_1 d^3y_2 & \rho({\bf y_1}) \rho({\bf y_2}) \cr
&\Bigl[
\Bigl({\pa^2 \over \pa y_1^i \pa y_1^j}-{1 \over 3}\delta_{ij}\triangle_1 \Bigr)
-3\Bigl({\pa^2 \over \pa y_1^i \pa y_2^j}-{1 \over 3}\delta_{ij}\triangle_{12} 
\Bigr) \Bigr] \ln(r_1+r_2+r_{12}), \cr}
\eqno\eq
$$
where we introduce 
$$
\eqalign{
\triangle_1=&{\pa^2 \over \pa y_1^k \pa y_1^k}, \cr
\triangle_{12}=&{\pa^2 \over \pa y_1^k \pa y_2^k}. \cr}
\eqno\eq
$$
Using relations $\Delta_{flat} |{\bf x}-{\bf y}|=2/ |{\bf x}-{\bf y}|$ and 
$\Delta_{flat}  |{\bf x}-{\bf y}|^3=12 |{\bf x}-{\bf y}|$,
$\four h_{ij}^{(C)}$, $\four h_{ij}^{(\rho)}$ and $\four h_{ij}^{(V)}$ are 
solved as 
$$
\four h_{ij}^{(C)}=
2{\pa \over \pa x^i}\int(\rho v^j)^{\cdot}\vert {\bf x}-{\bf y} \vert d^3y
+2{\pa \over \pa x^j}\int(\rho v^i)^{\cdot}\vert {\bf x}-{\bf y} \vert d^3y
+{4 \over 3}\delta_{ij}\int\ddot\rho \vert {\bf x}-{\bf y} \vert d^3y, 
\eqno\eq
$$
$$
\four h_{ij}^{(\rho)}=
{1 \over 12}{\pa^2 \over \pa x^i \pa x^j}\int\ddot\rho 
\vert {\bf x}-{\bf y} \vert^3 d^3y
-{1 \over 3}\delta_{ij}\int\ddot\rho \vert {\bf x}-{\bf y} \vert d^3y, 
{\hskip 4.8cm}\eqno\eq
$$ 
$$
\four h_{ij}^{(V)}=
{\pa^2 \over \pa x^i \pa x^j}\int\Bigl(\rho v^2+3P-{\rho U \over 2} \Bigr)
\vert {\bf x}-{\bf y} \vert d^3y 
-{2 \over 3}\delta_{ij}\int{\Bigl(\rho v^2+3P-\rho U / 2 \Bigr)
\over \vert {\bf x}-{\bf y} \vert} d^3y. 
\eqno\eq
$$
In total, we obtain
$$
\four h_{ij}=\four h_{ij}^{(S)}+\four h_{ij}^{(U)}
+\four h_{ij}^{(C)}+\four h_{ij}^{(\rho)}
+\four h_{ij}^{(V)}.
\eqn\hhheq
$$

\section{Treatment as a boundary value problem}

The above expression for $\four h_{ij}$ is quite interesting because 
it only consists of integrals over the matter. However, 
in actual numerical simulations, it will take a very 
long time to perform the direct integration. 
Therefore, we also propose other strategies where Eq.$\fourh$ is solved
as the boundary value problem.
Here, we would like to emphasize that the boundary condition should be
imposed at $r(=\vert {\bf x} \vert) \gg \vert {\bf y_1} \vert,
\vert {\bf y_2} \vert$, but $r$ does not have to be greater than 
$\lambda$,
where $\lambda$ is a typical wave length of gravitational waves. We only
need to impose $r > R$(a typical size of matter).
This means that we do not need a large amount of grid numbers compared with 
the case of fully general relativistic simulations, 
in which we require $r > \lambda \gg R$.

First of all, we consider the equation 
$$
\Delta_{flat}\Bigl(\four h^{(S)}_{ij}+\four h^{(U)}_{ij} \Bigr)=
\four \tau_{ij}^{(S)}+\four \tau_{ij}^{(U)}.\eqno\eq
$$
Since its source term behaves as $O(r^{-6})$ at $r \rightarrow 
\infty$, this equation can be accurately solved 
under the boundary condition at $r > R$ as 
$$\eqalign{
\four h_{ij}^{(S)}+\four h_{ij}^{(U)}&=
{2 \over r}\Bigl(\ddot I_{ij}-{1 \over 3}\delta_{ij}\ddot I_{kk}\Bigr)\cr
&+{2 \over 3r^2}\Bigl(n^k \ddot I_{ijk}-{1 \over 3}\delta_{ij}n^k \ddot I_{llk}
+2n^k(\dot S_{ikj}+\dot S_{jki})-{4 \over 3}\delta_{ij}n^k \dot S_{lkl} 
\Bigr)+O(r^{-3}),\cr}
\eqno\eq
$$
where
$$\eqalign{
&I_{ijk}=\int\rho x^ix^jx^k d^3x,\cr
&S_{ijk}=\int\rho (v^ix^j-v^jx^i)x^k d^3x.\cr}\eqno\eq
$$

Next, we consider the equations for $\four h_{ij}^{(C)}$, 
$\four h_{ij}^{(\rho)}$ and $\four h_{ij}^{(V)}$. Noting the identity, 
$$
\ddot \rho=-(\rho v^i)_{,i}^{\cdot}
=(\rho v^iv^j)_{,ij}+\Delta_{flat} P-(\rho U_{,i})_{,i}~, \eqno\eq
$$
we find the following relations;
$$\eqalign{
&\int \ddot \rho |{\bf x}-{\bf y}|d^3y=
-\int d^3y 
{x^i-y^i \over  \vert {\bf x}-{\bf y} \vert}(\rho v^i)^{\cdot},\cr
&\int \ddot \rho |{\bf x}-{\bf y}|^3d^3y=
3\int d^3y\biggl[ \rho v^i v^j{(x^i-y^i)(x^j-y^j) \over  |{\bf x}-{\bf y}|}
+\Bigl(4 P+\rho v^2
-\rho U_{,i} (x^i-y^i)\Bigr) |{\bf x}-{\bf y}|\biggr].\cr}
\eqn\rereeq
$$
Using Eqs.$\rereeq$, 
$\four h_{ij}^{(C)}$, $\four h_{ij}^{(\rho)}$ and 
$\four h_{ij}^{(V)}$ in Eqs.(5.16-18) can be rewritten as 
$$
\four h_{ij}^{(C)}=
2\int(\rho v^j)^{\cdot}{x^i-y^i \over \vert {\bf x}-{\bf y} \vert }d^3y
+2\int(\rho v^i)^{\cdot}{x^j-y^j \over \vert {\bf x}-{\bf y} \vert}d^3y 
-{4 \over 3}\delta_{ij}\int 
(\rho v^k)^{\cdot} {x^k-y^k \over  \vert {\bf x}-{\bf y} \vert} d^3y,
\eqno\eq
$$
$$\eqalign{
\four h_{ij}^{(\rho)}&=
{1 \over 4}{\pa^2 \over \pa x^i \pa x^j}\int 
\rho v^k v^l {(x^k-y^k)(x^l-y^l) \over  |{\bf x}-{\bf y}| } d^3y
+{1 \over 3}\delta_{ij}\int
(\rho v^k)^{\cdot} {x^k-y^k \over  \vert {\bf x}-{\bf y} \vert} d^3y \cr
&~~~+{1 \over 2}\biggl\{
{\pa \over \pa x^i }\int P' { (x^j-y^j) \over  |{\bf x}-{\bf y}| }d^3y 
+{\pa \over \pa x^j }\int P'{ (x^i-y^i) \over  |{\bf x}-{\bf y}| }d^3y\biggr\}
\cr
&~~~-{1 \over 8}\biggl\{
2\int \rho { U_{,j}(x^i-y^i)+U_{,i}(x^j-y^j) \over 
|{\bf x}-{\bf y}| } d^3y \cr
&~~~~~~~~~~~~~+x^k{\pa \over \pa x^i }\int \rho {U_{,k}(x^j-y^j) \over
|{\bf x}-{\bf y}|}  d^3y
+x^k{\pa \over \pa x^j }\int \rho {U_{,k}(x^i-y^i) \over
|{\bf x}-{\bf y}|} d^3y \biggr\},\cr}\eqno\eq
$$
and
$$\eqalign{
\four h_{ij}^{(V)}&=
{1 \over 2}\Bigl[
{\pa \over \pa x^i}\int\Bigl(\rho v^2+3P-{\rho U \over 2} \Bigr)
{x^j-y^j \over \vert {\bf x}-{\bf y} \vert} d^3y
+{\pa \over \pa x^j}\int\Bigl(\rho v^2+3P-{\rho U \over 2} \Bigr)
{x^i-y^i \over \vert {\bf x}-{\bf y} \vert} d^3y \Bigr] \cr
&~~~~~~~~-{2 \over 3}\delta_{ij}\int{\Bigl(\rho v^2+3P-\rho U / 2 \Bigr)
\over \vert {\bf x}-{\bf y} \vert } d^3y,\cr}
\eqno\eq
$$
where $P'=P+\rho v^2/4+\rho U_{,l}y^l/4$. 
From the above relations, 
$\four h_{ij}^{(C)}$, $\four h_{ij}^{(\rho)}$ and $\four h_{ij}^{(V)}$
become 
$$\eqalign{
\four h_{ij}^{(C)}&=2(x^i \three \dot P^j+x^j \three \dot P^i-Q_{ij})
+{4 \over 3}\delta_{ij}\Bigl({Q_{kk} \over 2}-x^k \three \dot P^k \Bigr),
\cr
\four h_{ij}^{(\rho)}&={1 \over 4}{\pa^2 \over \pa x^i\pa x^j}\Bigl(
V_{kl}^{(\rho v)}x^kx^l-2V_k^{(\rho v)} x^k+V^{(\rho v)} \Bigr)
+{1 \over 3}\delta_{ij} \Bigl(x^k \three \dot P_k-{Q_{kk}\over 2}\Bigr) \cr
&~~~~~+{1 \over 2}\Bigl\{{\pa \over \pa x^i}\Bigl(V^{(P)}x^j-V^{(P)}_j \Bigr)
+{\pa \over \pa x^j}\Bigl(V^{(P)}x^i-V^{(P)}_i \Bigr) \Bigr\}\cr
&~~~~~-{1 \over 8}\Bigl\{ 2\Bigl(x^iV^{(\rho U)}_j+x^iV^{(\rho U)}_i
-V^{(\rho U)}_{ij}-V^{(\rho U)}_{ji}\Bigr) \cr
&~~~~~~~~~~+x^k{\pa \over \pa x^i}\Bigl(x^j V^{(\rho U)}_k -V^{(\rho U)}_{kj}
\Bigr)
+x^k{\pa \over \pa x^j}\Bigl(x^i V^{(\rho U)}_k -V^{(\rho U)}_{ki}\Bigr)
\Bigr\}, 
\cr
\four h_{ij}^{(V)}&={1 \over 2}\Bigl(Q^{(I)}_{,j}x^i+Q^{(I)}_{,i}x^j-
Q^{(I)}_{i,j}-Q^{(I)}_{j,i}\Bigr)+{1 \over 3}Q^{(I)} \delta_{ij},\cr}\eqno\eq
$$
where 
$$\eqalign{
&\Delta_{flat}\three P_i=-4\pi \rho v^i,\cr
&\Delta_{flat}Q_{ij}=-4\pi\Bigl\{x^j (\rho v^i)^{\cdot}
                                +x^i (\rho v^j)^{\cdot}\Bigr\},\cr
&\Delta_{flat} Q^{(I)}=-4\pi \Bigl(\rho v^2+3P-{1 \over 2}\rho U\Bigr),\cr
&\Delta_{flat} Q^{(I)}_i
     =-4\pi \Bigl(\rho v^2+3P-{1 \over 2}\rho U \Bigr)x^i,\cr
&\Delta_{flat} V_{ij}^{(\rho v)}=-4\pi \rho v^i v^j,\cr
&\Delta_{flat} V_i^{(\rho v)}=-4\pi \rho v^i v^j x^j,\cr
&\Delta_{flat} V^{(\rho v)}=-4\pi \rho (v^j x^j)^2,\cr
&\Delta_{flat} V^{(P)}=-4\pi P',\cr
&\Delta_{flat} V^{(P)}_i=-4\pi P' x^i ,\cr
&\Delta_{flat} V^{(\rho U)}_i=-4\pi\rho U_{,i},\cr
&\Delta_{flat} V^{(\rho U)}_{ij}=-4\pi\rho U_{,i}x^j.\cr}\eqn\poiseq
$$
Therefore, $\four h_{ij}^{(C)}$, $\four h_{ij}^{(\rho)}$ and 
$\four h_{ij}^{(V)}$ can be derived from the above potentials which 
satisfy the Poisson equations with compact sources.

We note that instead of the above procedure, 
we may solve the Poisson equation for $\four h_{ij}$ 
carefully imposing the boundary condition for $r \gg R$ as 
$$
\eqalign{
\four h_{ij}=&
{1 \over r}
\Bigl\{ {1 \over 4}I_{ij}^{(2)}+{3 \over 4}n^k\Bigl(
n^iI_{kj}^{(2)}+n^jI_{ki}^{(2)}\Bigr) \cr
&~~~~~~~-{5 \over 8}n^in^jI_{kk}^{(2)}+{3 \over 8}n^in^jn^kn^lI_{kl}^{(2)} 
+{1 \over 8}\delta_{ij}I_{kk}^{(2)}-{5 \over 8}\delta_{ij}n^kn^lI_{kl}^{(2)} 
\Bigr\} \cr
&+{1 \over r^2}
\Bigl[ \Bigl\{
-{5 \over 12}n^kI_{ijk}^{(2)}-{1\over 24}(n^iI_{jkk}^{(2)}+n^jI_{ikk}^{(2)})
+{5 \over 8}n^kn^l(n^iI_{jkl}^{(2)}+n^jI_{ikl}^{(2)}) \cr
&~~~~~~~-{7 \over 8}n^in^jn^kI_{kll}^{(2)}+{5 \over 8}n^in^jn^kn^ln^m
I_{klm}^{(2)}+{11 \over 24}\delta_{ij}n^kI_{kll}^{(2)}
-{5 \over 8}\delta_{ij}n^kn^ln^mI_{klm}^{(2)} 
\Bigr\} \cr
&~~~~~~~+\Bigl\{
{2 \over 3}n^k(\dot S_{ikj}+\dot S_{jki})-{4 \over 3}(n^i\dot S_{jkk} 
+n^j\dot S_{ikk}) \cr
&~~~~~~~+2n^kn^l(n^i\dot S_{jkl}+n^j\dot S_{ikl})+2n^in^jn^k\dot S_{kll} 
+{2 \over 3}\delta_{ij}n^k\dot S_{kll} \Bigr\} \Bigr]+O(r^{-3}). \cr}
\eqn\bc
$$
It is verified that $O(r^{-1})$ and $O(r^{-2})$ parts 
satisfy the traceless and divergence-free conditions respectively. 
It should be noticed that $\four h_{ij}$ obtained in this way 
becomes meaningless at the far zone because Eq.$\fourh$, from which 
$\four h_{ij}$ is derived, is valid only in the near zone. 

\chapter{Conserved quantities}

The conserved quantities are gauge-invariant so that, 
in general relativity, they play important roles because we are able to 
compare various systems described in different gauge conditions using them. 
From the practical view, these are also useful for checking 
the numerical accuracy in simulations. Thus, in this section, we 
show several conserved quantities in the 2PN approximation. 

\section{Conserved Mass And Energy}

In general relativity, 
the volume integral of the mass density $\rho$ does not conserve, and 
instead we have the following conserved mass; 
$$
M_{\ast}=\int \rho_{\ast}d^3x. 
\eqno\eq
$$
In the PN approximation, $\rho_{\ast}$ is expanded as
$$
\eqalign{
\rho_{\ast}=\rho \Bigl[&1+\Bigl({1 \over 2}v^2+3 U\Bigr) \cr
&+\Bigl({3 \over 8}v^4+{7 \over 2}v^2 U
+{15 \over 4}U^2+6 \four\psi+\three\beta_i v^i\Bigr)+\six\delta_{\ast}+
O(c^{-7})\Bigr], }
\eqno\eq
$$
where $\six\delta_{\ast}$ denotes the 3PN contribution to $\rho_{\ast}$. 

Then, we consider the ADM mass which is also the conserved quantity. 
Since the asymptotic behavior of the conformal factor becomes 
$$
\psi=1+{M_{ADM} \over 2 r}+O\Bigl({1 \over r^2}\Bigr), 
\eqno\eq
$$
the ADM mass in the PN approximation becomes 
$$
\eqalign{
M_{ADM}&=-{1 \over 2\pi} \int \Delta_{flat} \psi d^3x \cr
&=\int d^3x \rho \Bigl[\Bigl\{1+\Bigl(v^2+\varepsilon+{5 \over 2}U
\Bigr) 
+\Bigl(v^4+{13 \over 2}v^2 U+v^2\varepsilon+{P \over \rho}v^2
+{5 \over 2}U\varepsilon+{5 \over 2}U^2 \cr
&~~~~~+5\four\psi+2\three\beta_i v^i\Bigr)\Bigr\} 
+{1 \over 16\pi\rho}\Bigl(\three\tilde A_{ij} \three\tilde A_{ij}
-{2 \over 3}\three K^2 \Bigr)+\six \delta_{ADM}+O(c^{-7}) \Bigr], \cr}
\eqno\eq
$$
where $\six \delta_{ADM}$ denotes the 3PN contribution. 

Using these two conserved quantities, we can define 
the conserved energy as follows; 
$$
\eqalign{
E\equiv &M_{ADM}-M_{\ast} \cr
=&\int d^3x \rho \Bigl[\Bigl\{\Bigl({1 \over 2}v^2+\varepsilon
-{1 \over 2}U\Bigr) \cr
&~~~~~~+\Bigl({5 \over 8}v^4+3v^2 U+v^2\varepsilon+{P \over \rho}v^2
+{5 \over 2}U\varepsilon-{5 \over 4}U^2-\four\psi+\three\beta_i v^i
\Bigr)\Bigr\} \cr
&~~~~~~+{1 \over 16\pi\rho}\Bigl(\three\tilde A_{ij} 
\three\tilde A_{ij}-{2 \over 3}\three K^2\Bigr)
+\Bigl(\six\delta_{ADM}-\six\delta_{\ast}\Bigr)+O(c^{-7})\Bigr] \cr
\equiv & E_{N}+E_{1PN}+E_{2PN}+\cdots . \cr}
\eqno\eq
$$
We should notice that the following equation holds
$$
\int\three\tilde A_{ij} \three\tilde A_{ij} d^3x=
-8\pi\int\rho v^i \three\beta_i d^3x+\int\Bigl({2 \over 3}\three K^2
+2 \dot U \three K\Bigr)d^3x ,
\eqno\eq
$$
where we use the identities derived from Eqs.$\dotu$ and $\betth$ 
$$
\eqalign{
\int \three\beta_{i,j}\three\beta_{i,j}d^3x
&=-16\pi\int\rho v^i \three\beta_i d^3x
+\int(\three K^2+2\dot U\three K-3\dot U^2)d^3x, \cr
\int \three\beta_{i,j}\three\beta_{j,i}d^3x
&=\int(\three K^2+6\dot U \three K+9\dot U^2)d^3x .}
\eqno\eq
$$
Using these relations, we obtain the Newtonian and the first PN energies as 
$$
E_{N}=\int\rho\Bigl({1 \over 2}v^2+\varepsilon-{1 \over 2}U\Bigr)d^3x ,
\eqno\eq
$$
and
$$
E_{1PN}=\int d^3x\Bigl[\rho\Bigl({5 \over 8}v^4+{5 \over 2}v^2 U
+v^2\varepsilon+{P \over \rho}v^2+2 U\varepsilon-{5 \over 2}U^2
+{1 \over 2}\three\beta_i v^i 
\Bigr)+{1 \over 8 \pi}\dot U\three K \Bigr]. 
\eqno\eq
$$
$E_{1PN}$ can be rewritten immediately in the following form 
used by Chandrasekhar\refmark\chandr; 
$$
E_{1PN}=\int d^3x\rho \Bigl[{5 \over 8}v^4+{5 \over 2}v^2 U+v^2 
\Bigl(\varepsilon+{P \over \rho}\Bigr)+2 U\varepsilon-{5 \over 2}U^2
-{1 \over 2}v^i q_i \Bigr] , 
\eqno\eq
$$
where $q_i$ is the first PN shift vector in the standard PN gauge(\ie, 
$\three K=0$) and satisfies 
$$
\Delta_{flat} q_i=-16\pi\rho v^i+\dot U_{,i} .
\eqno\eq
$$

$E_{2PN}$ is calculated from the 3PN quantities $\six\delta_{\ast}$ and 
$\six\delta_{ADM}$. $\six\delta_{\ast}$ becomes 
$$
\eqalign{
\six\delta_{\ast}=&
{5 \over 16}v^6+{33 \over 8}v^4 U+v^2 \Bigl( 5\four\psi
+{93 \over 8}U^2+{3 \over 2}\three\beta_i v^i-X \Bigr) \cr
&~~+6\six\psi+15U\four\psi+{5 \over 2}U^3+7\three\beta_i v^i U
+{1 \over 2}\four h_{ij}v^iv^j
+{1 \over 2}\three\beta_i\three\beta_i+\five\beta_iv^i , \cr}
\eqno\eq
$$
and we obtain
$$
\six M_{\ast}=\int \rho \six\delta_{\ast} d^3x.
\eqno\eq
$$ 
The Hamiltonian constraint at $O(c^{-8})$ becomes 
$$
\eqalign{
&\Delta_{flat}\eight\psi-\four h_{ij}\four\psi_{,ij}
-{1 \over 2}\six h_{ij}U_{,ij} \cr
&=-{1 \over 32}(2\four h_{kl,m}\four h_{km,l}
+\four h_{kl,m}\four h_{kl,m}) \cr
&~~~-2\pi\six\rho_{\psi}
-{1 \over 4} \Bigl( \three\tilde A_{ij}\five\tilde A_{ij}
-{2 \over 3}\three K\five K \Bigr)
-{1 \over 16}U \Bigl( \three\tilde A_{ij}\three\tilde A_{ij}
-{2 \over 3}\three K^2 \Bigr), }
\eqno\eq
$$
where we define $\six\rho_{\psi}$ as 
$$
\eqalign{
\six\rho_{\psi}=&\rho \Bigl[ v^6+v^4 \Bigl( \varepsilon+{P \over \rho}
+{21 \over 2}U \Bigr) +v^2 \Bigl\{ {13 \over 2}U \Bigl( \varepsilon+
{P \over \rho} \Bigr) +9\four\psi-2X+20U^2 \Bigr\} \cr
&~~~~~+\varepsilon \Bigl( 5\four\psi+{5 \over 2}U^2 \Bigr)+
5\six\psi+10U\four\psi+{5 \over 4}U^3 \cr
&~~~~~+\four h_{ij}v^iv^j+2\three\beta_iv^i \Bigl\{ 2v^2+\varepsilon
+{P \over \rho}+{13 \over 2}U \Bigr\} +2\five\beta_iv^i
+\three\beta_i\three\beta_i \Bigr]. }
\eqno\eq
$$
Making use of relations $\four h_{ij,j}=0$ and 
$\six h_{ij,j}=0$, we obtain 
$$
\eqalign{
\six M_{ADM}=&\int d^3x \six\rho_{\psi} \cr
&+{1 \over 8\pi}\int d^3y 
\Bigl( \three\tilde A_{ij}\five\tilde A_{ij}
-{2 \over 3}\three K\five K \Bigr) \cr
&+{1 \over 32\pi}\int d^3y 
U \Bigl( \three\tilde A_{ij}\three\tilde A_{ij}
-{2 \over 3}\three K^2 \Bigr), }
\eqno\eq
$$
where we assume $\six h_{ij} \to O(r^{-1})$ as $r \to \infty$. 
Although this assumption must be verified by performing the 3PN expansions 
which have not been done here, it seems reasonable in the asymptotically 
flat spacetime. From $\six M_{ADM}$ and $\six M_{\ast}$, 
we obtain the conserved energy at the 2PN order
$$
\eqalign{
E_{2PN}=&\six M_{ADM}-\six M_{\ast} \cr
=&\int d^3x \rho \Bigl[ {11 \over 16}v^6+v^4 \Bigl( \varepsilon
+{P \over \rho}+{51 \over 8}U \Bigr) \cr 
&~~~~~~~~~~~~+v^2 \Bigl\{ 4\four\psi-X
+{13 \over 2}U \Bigl( \varepsilon+{P \over \rho} \Bigr) +{67 \over 8}U^2
+{5 \over 2}\three\beta_iv^i \Bigr\} \cr
&~~~~~~~~~~~~+\varepsilon \Bigl( 5\four\psi+{5 \over 2}U^2 \Bigr)
-\six\psi-5U\four\psi-{5 \over 4}U^3  \cr
&~~~~~~~~~~~~+{1 \over 2}\four h_{ij}v^iv^j+2\three\beta_iv^i \Bigl( 
\varepsilon+{P \over \rho}+3U \Bigr) +\five\beta_iv^i
+{ 1\over 2}\three\beta_i\three\beta_i \Bigr\} \Bigr] \cr
&+{1 \over 8\pi}\int d^3y 
\Bigl( \three\tilde A_{ij}\five\tilde A_{ij}
-{2 \over 3}\three K\five K \Bigr) \cr
&+{1 \over 32\pi}\int d^3y 
U \Bigl( \three\tilde A_{ij}\three\tilde A_{ij}
-{2 \over 3}\three K^2 \Bigr) . }
\eqno\eq
$$
When we use the relation, $\int d^3x \rho\six\psi=-{1 \over 4\pi}
\int d^3x U\Delta\six\psi$, we obtain 
$$
\eqalign{
E_{2PN}=&\int d^3x \rho \Bigl[ {11 \over 16}v^6+v^4 \Bigl( \varepsilon
+{P \over \rho}+{47 \over 8}U \Bigr) \cr 
&~~~~~~~~~~~~+v^2 \Bigl\{ 4\four\psi-X
+6U \Bigl( \varepsilon+{P \over \rho} \Bigr) +{41 \over 8}U^2
+{5 \over 2}\three\beta_iv^i \Bigr\} \cr
&~~~~~~~~~~~~+\varepsilon \Bigl( 5\four\psi+{5 \over 4}U^2 \Bigr)
-{15 \over 2}U\four\psi-{5 \over 2}U^3  \cr
&~~~~~~~~~~~~+{1 \over 2}\four h_{ij}v^iv^j+2\three\beta_iv^i \Bigl\{ 
\Bigl( \varepsilon+{P \over \rho} \Bigr) +5U \Bigr\} +\five\beta_iv^i
+{ 1\over 2}\three\beta_i\three\beta_i \Bigr] \cr
&+{1 \over 8\pi}\int d^3y 
\Bigl( \four h_{ij}UU_{,ij}+\three\tilde A_{ij}\five\tilde A_{ij}
-{2 \over 3}\three K\five K \Bigr) . }
\eqno\eq
$$

\section{Conserved linear momentum}

When we use the center of mass system as usual, 
the linear momentum of the system should vanish.
However, it may arise from numerical errors 
in numerical calculations.
Since it is useful for investigation of the numerical accuracy, 
we mention the linear momentum derived from 
$$
\eqalign{
P_i=&{1 \over 8\pi}\lim_{r \to \infty}\oint
\Bigl( K_{ij}n^j-K n^i \Bigr) dS \cr
=&{1 \over 8\pi}\lim_{r \to \infty}\oint
\Bigl( \psi^4 \tilde A_{ij} n^j-{2 \over 3}K n^i \Bigr) dS ,}
\eqno\eq
$$
where the surface integrals are taken over a sphere of constant $r$.
Since the asymptotic behavior of $\tilde A_{ij}$ is determined by
$$
\three\tilde A_{ij}={1 \over 2}\bigl(\three\beta_{i,j}+\three\beta_{j,i}
-{2 \over 3}\delta_{ij}\three\beta_{l,l}\Bigr)+O(r^{-3}) ,
\eqno\eq
$$
and
$$
\five\tilde A_{ij}={1 \over 2}\Bigl(\five\beta_{i,j}+\five\beta_{j,i}
-{2 \over 3}\delta_{ij}\five\beta_{l,l}\Bigr)+\five\tilde A_{ij}^{TT}
+O(r^{-3}) ,
\eqno\eq
$$
the leading term of the shift vector is necessary.
Using the asymptotic behavior
$$
\three\beta_i=-{7 \over 2}{l_i \over r}
-{1 \over 2}{n^i n^j l_j \over r}+O(r^{-2}) ,
\eqno\eq
$$
the following relation is obtained 
$$
\int\Bigl(\three\beta_{i,j}+\three\beta_{j,i}
-{2 \over 3}\delta_{ij}\three\beta_{l,l}\Bigr) n^j dS
=16\pi l_i ,
\eqno\eq
$$
where $l_i=\int \rho v^i d^3x$.
Therefore the Newtonian linear momentum is 
$$
P_N\,^i=\int d^3x \rho v^i .
\eqno\eq
$$
Similarly the first PN linear momentum is obtained as follows; 
$$
P_{1PN}\,^i=\int d^3x \rho \Bigl[v^i\Bigl(v^2+\varepsilon+6U+{P \over \rho}
\Bigr)+\three\beta_i\Bigr].
\eqno\eq
$$
$P_{2PN}\,^i$ is obtained by the similar procedure. 

\chapter{Summary}

In this paper, we 
have developed the PN approximation in the (3+1) formalism of general 
relativity. In this formalism, it is clarified what kind of 
gauge condition is suitable for each problem such as how to extract 
the waveforms of gravitational waves and how to describe equilibrium 
configurations.  
It was found that the combination of the conformal slice and the 
transverse gauge is useful to separate  the wave part and the non-wave part 
in the metric variables such as  $h_{ij}$ and $\psi$. 
We also found that, in order 
to describe the equilibrium configuration, the conformal 
slice is not useful and instead we had better use the maximal slice. 
Although we restricted ourselves within some gauge conditions in this paper, 
we may try to use any gauge condition and investigate its property 
relatively easily in the (3+1) formalism, compared with in the standard PN 
approximation performed so far\refmark\chandra. 

We have also developed a formalism for the hydrodynamic equation 
accurate up to 2.5PN order. For the sake of an actual 
numerical simulation, we carefully consider methods to solve the various 
metric quantities, especially, the 2PN tensor potential $\four h_{ij}$. 
We found it possible to solve them by using 
standard numerical methods. 
Thus, the formalism developed in this paper will be useful also in numerical 
calculations. 

Here, we would like to emphasize that from the 2PN order, the 
tensor part of the 3-metric, $\tilde\gamma_{ij}$, cannot be neglected  
even if we ignore gravitational waves. 
Recently, Wilson and Mathews\refmark\wilson 
presented numerical equilibrium configurations of binary neutron stars 
using a semi-relativistic approximation, in which 
they assume the spatially conformal flat metric as the spatial 3-metric, 
\ie, $\tilde \gamma_{ij}=\delta_{ij}$. 
Thus, in their method, a 2PN term, $h_{ij}$, was completely neglected. 
This means that their results unavoidably have an error of the 2PN order 
which will become $\sim (M/R)^2 \sim 1-10\%$. If we hope to obtain 
a general relativistic eqiulibrium configuration of binary neutron stars 
with a better accuracy(say less than $1\%$), 
we should take into account the tensor part of the 3-metric. 

In section 3,  we used several slice conditions and 
investigated their properties, but, as for 
the spatial gauge condition, we fix it to the transverse gauge for the 
sake of convenience. It is not clear, however,  whether this is the best 
gauge condition in numerical relativity. 
In numerical relativity, the shift vector 
plays a very important role to reduce the coordinate shear. If we fail to 
choose the appropriate condition, the coordinate shear in the spatial metric 
will continue to grow, and as a result, the simulation will break down. 
The minimal distortion gauge which was proposed by 
Smarr and York\refmark\smarr is a candidate which may efficiently reduce 
the coordinate shear. 
Even if we use this gauge condition in the 
PN analysis, equations for $\four h_{ij}$, $\five h_{ij}$, 
$\three \beta_i$, $\five\beta_i$ and $\six\beta_i$ remain unchanged, 
but higher order terms of $h_{ij}$ and $\beta_i$ may slightly 
change. If we investigate the effects due to the difference, 
we may be able to give some important suggestions about the 
gauge condition appropriate for numerical relativity. 

In this paper, we mainly payed attention to the PN hydrodynamic equation in 
the (3+1) formalism to describe the final hydrodynamic 
merging phase of coalescing 
binary neutron stars. This formalism, however, may be useful 
also for gravitational wave generation problems 
in the inspiraling phase, which is 
usually performed using the harmonic gauge\refmark{\will, \blan}. 
Since we can choose 
any gauge conditions in the (3+1) formalism, it may be possible to reduce 
some lengthy calculations which appear in previous works using the harmonic 
gauge\refmark{\will, \blan}. 
This problem is interesting and important as a future work. 

\vskip 5mm

\centerline{\fourteenrm Acknowledgment}

We thank T. Nakamura for his suggestion to pursue this problem 
and useful discussions. We also thank M. Sasaki, T. Tanaka and K. Nakao 
for helpful discussion. 
H. A. thanks Prof. S. Ikeuchi and Prof. M. Sasaki for encouragements. 
This work was in part supported by the Japanese Grant-in-Aid for 
Scientific Research of the Ministry of Education, Science, and Culture, 
No. 07740355.

\Appendix{A}
\centerline
{\fourteenrm Calculation of $\five h_{ij}$}

We make use of the transverse property of $\tau_{ij}$, which 
is guaranteed by the transverse gauge condition, in order to obtain 
$\five h_{ij}$. Using the following identity
$$
\four\tau_{ij}=(\four\tau_{ik}x^j)_{,k},
\eqno\eq
$$
Eq.$\quadr$ can be rewritten in the surface integral form
$$
\five h_{ij}={1 \over 4\pi}{\pa \over \pa t}
\int \four\tau_{ik}x^j n^k dS. 
\eqno\eq
$$
Thus, we only need to estimate terms of  $O(r^{-3})$ in $\four\tau_{ij}$, 
which come only from the shift vector in the conformal slice as 
$$
\three\beta_i={1 \over r^2}\Bigl(n^j Z_{ij}+{1 \over 2}n^j \dot I_{ij}
+{1 \over 4}n^i \dot I_{kk}-{3 \over 4}n^i n^j n^k \dot I_{jk}
+{n^i \over 4\pi}\int\three K d^3x \Bigr)+O(r^{-3}),\eqno\eq
$$
where
$$
Z_{ij}(t)=-4\int\rho v^i y^j d^3x.\eqno\eq
$$
Here, note the following relations as 
$$
{\pa \over \pa t}Z_{ij}=-2 \ddot I_{ij}, 
\eqno\eq
$$
and 
$$
\int \three \dot K d^3x =2\pi \ddot I_{kk}.\eqno\eq
$$
Therefore, the relevant terms of $\four \tau_{ij}$ for the surface integral 
become 
$$
\eqalign{
\four \tau_{ij} \rightarrow 
&\three\dot\beta_{i,j}+\three\dot\beta_{j,i}-{2 \over 3}\delta_{ij} 
\three\dot\beta_{l,l} \cr
&={1 \over r^3} \Bigl[ \Bigl\{ -3\ddot I_{ij}+3 \Bigl( \ddot I_{ik}n^kn^j
+\ddot I_{jk}n^kn^i \Bigr) -{9 \over 2}\ddot I_{kk} n^in^j
+{15 \over 2}\ddot I_{kl}n^in^jn^kn^l \Bigr\} \cr
&~~~~~-{1 \over 3}\delta_{ij} \Bigl\{ -{15 \over 2}\ddot I_{kk}+{27 \over 2}
\ddot I_{kl}n^kn^l \Bigr\} \Bigr]+O(r^{-4}). \cr}
\eqno\eq
$$
Thus we obtain $h_{ij}$ at the 2.5PN order,
$$
\eqalign{
\five h_{ij}&=-{1 \over 4\pi}{\pa \over \pa t}
\int \Bigl(\three\dot\beta_{i,k}+\three\dot\beta_{k,i}-{2 \over 3}\delta_{ik}
\three\dot\beta_{l,l}\Bigr)x^j n^k dS \cr
&=-{4 \over 5}\bI_{ij}^{(3)}(t). \cr}
\eqno\eq
$$
This derivation seems fairly simple owing to the gauge condition.
Thus, it is expected that higher order calculations, 
say at 3.5PN order, may become easier in this gauge condition.

\Appendix{B}

In this appendix, we briefly comment on a method to
solve the Poisson equations
for $\three\beta_i$ and $\five\beta_i$, \ie, Eqs.(2.51) and (2.53).
Since the source terms of them
have terms such as $-\dot U_{,i}$ 
and $-2\four \dot \psi_{,i}$ which behaves as $O(r^{-2})$ 
at $r \rightarrow \infty$\footnote{\dagger}{Note that in the Newtonian 
limit, $-\dot U_{,i}$ is $O(r^{-4})$, but at the 1PN order, 
it becomes $O(r^{-2})$ because $\int\dot\rho dV\neq 0$.}, it seems that
a technical problem arises in solving these equations in numerical 
calculation.
However, this is easily overcome in a simple manner.
We consider the case of the maximal slice for simplicity, but 
other cases may be treated similarly. 

First of all, we write $\three\beta_i$ and $\five\beta_i$ as,
$$\eqalign{
\three \beta_i &=-4\three P_i+{1 \over 2}\dot \chi_{,i},\cr
\five \beta_i  &=-4\five  P_i+{1 \over 2}\four \dot \chi_{,i},\cr}
\eqno\eq
$$
where $\chi$ and $\three P_i$ satisfy Eq.(4.4) and the first equation of 
Eqs.$\poiseq$, respectively. $\five P_i$ and
$\four\chi$ satisfy the following Poisson equations;  
$$\eqalign{
\Delta_{flat} \five P_i &=-4\pi \rho \Bigl[v^i\Bigl(v^2+2U+\varepsilon+
{P \over \rho} \Bigr)+\three\beta_i \Bigr]+2U_{,j} \three \tilde A_{ij}
-{1 \over 8}(\dot U U)_{,i}-{1 \over 4}(\three\beta_l U_{,l} )_{,i}~,\cr
\Delta_{flat}\four\chi &=-4\four\psi.\cr}\eqno\eq
$$
$\four\chi$ can be written as 
$$
\four\chi=-\int \rho_4|{\bf x}-{\bf y}|d^3y,\eqn\chieq
$$
where 
$$
\rho_4=\rho\Bigl(v^2+\varepsilon+{5 \over 2}U \Bigr).\eqno\eq
$$
From Eqs.(4.4) and $\chieq$, $\chi_{,i}$ and $\four\chi_{,i}$ become 
$$\eqalign{
\chi_{,i}&=-\int d^3y \rho {x^i-y^i \over |{\bf x}-{\bf y}|}
=-x^iU+\eta_i,\cr
\four \chi_{,i}&=-\int d^3y \rho_4 {x^i-y^i \over |{\bf x}-{\bf y}| }
=-2x^i \four\psi+\four \eta_i,\cr}\eqno\eq
$$
where $ \eta_i$ and $\four\eta_{i}$ satisfy
$$\eqalign{
\Delta_{flat} \eta_i&=-4\pi\rho x^i,\cr
\Delta_{flat} \four \eta_i&=-4\pi\rho_4 x^i.\cr}\eqno\eq
$$
Hence, 
$$\eqalign{
\three\beta_i&=-4\three P_i-{1 \over 2}\Bigl(x^i \dot U-\dot \eta_{i}
\Bigr),\cr
\five\beta_i&=-4\five P_i-{1 \over 2}\Bigl(2x^i \four\dot \psi-\four
\dot \eta_{i}\Bigr).\cr}\eqno\eq
$$
Since the source terms of the Poisson equations for $\three P_i$,
$\five P_i$, $\eta_i$ and $\four\eta_i$ behaves as
$O(r^{-n})$, where $n\geq 5$, at $r \rightarrow \infty$\footnote{\star}{
Note that the non-compact sources of the Poisson equation for 
$\five P_i$ may be regarded as $O(r^{-5})$ in the 2PN approximation because 
$\dot U$ is $O(r^{-3})$ in the Newtonian order.},
these vector potentials can be accurately 
obtained by solving the Poisson equations for them 
under appropriate boundary conditions.
Thus, there is no difficulty to obtain $\three\beta_i$ and $\five \beta_i$.

Finally, we note that the above method is not unique prescription. 
For example, $\three \beta_i$ in the first PN approximation\footnote{\ddagger}
{See also ref.[\bds].} may be expressed as
$$
\three\beta_i=-4\three P_i+{1 \over 2}\Bigl\{(\three P_k x^k)_{,i}-\chi_{2,i}
\Bigr\},\eqno\eq
$$
where $\chi_2$ satisfies 
$$
\Delta_{flat} \chi_2=-4\pi\rho v^i x^i.\eqno\eq
$$

\refout
\end